\documentclass[journal,10pt]{IEEEtranTCOM}

\normalsize

\usepackage[T1]{fontenc}
\usepackage{amsfonts}
\usepackage{amsmath}
\usepackage{amsthm}
\usepackage{amssymb}
\usepackage{caption}
\usepackage[labelformat=simple]{subcaption}
\usepackage{cite,graphicx,algorithm}
\usepackage{algorithmic}
\usepackage{color}
\usepackage{bm}

\ifCLASSINFOpdf
\else
\fi

\theoremstyle{plain}

\begin{document}
%
\title{OTFS Signaling for SCMA With Coordinated Multi-Point Vehicle Communications}
%
%
%

\author{Yao~Ge,~\IEEEmembership{Member,~IEEE,}
        Qinwen~Deng,~\IEEEmembership{Student Member,~IEEE,}
        David~Gonz\'{a}lez~G.,~\IEEEmembership{Senior Member,~IEEE,}
        Yong~Liang~Guan,~\IEEEmembership{Senior Member,~IEEE,}
        and~Zhi~Ding,~\IEEEmembership{Fellow,~IEEE}
\thanks{This study is supported under the RIE2020 Industry Alignment Fund—Industry Collaboration Projects (IAF-ICP) Funding Initiative, as well as cash and in-kind contribution from the industry partner(s). The work of Qinwen Deng and Zhi Ding was supported in part by the National Science Foundation under Grant 2029027.}
\thanks{Yao Ge and Yong Liang Guan are with the Continental-NTU Corporate Lab, Nanyang Technological University, Singapore (e-mail: yao.ge@ntu.edu.sg; eylguan@ntu.edu.sg). Yong Liang Guan is also with the School of Electrical and Electronics Engineering, Nanyang Technological University, Singapore.}
\thanks{Qinwen Deng and Zhi Ding are with the Department of Electrical and Computer Engineering, University of California at Davis, Davis, CA 95616 USA (e-mail: mrdeng@ucdavis.edu; zding@ucdavis.edu).}
\thanks{David~Gonz\'{a}lez~G. is with the Wireless Communications Technologies Group, Continental AG, Germany (e-mail: david.gonzalez.g@ieee.org).}}

%
%

\markboth{}%
{}
%



\maketitle

\begin{abstract}
This paper investigates an uplink coordinated multi-point (CoMP) coverage scenario, in which multiple mobile users are grouped for sparse code multiple access (SCMA), and served by the remote radio head (RRH) in front of them and the RRH behind them simultaneously. We apply orthogonal time frequency space (OTFS) modulation for each user to exploit the degrees of freedom arising from both the delay and Doppler domains. 
As the signals received by the RRHs in front of and behind the users experience respectively positive and negative Doppler frequency shifts, our proposed OTFS-based SCMA (OBSCMA) with CoMP system can effectively harvest extra Doppler and spatial diversity for better performance. Based on maximum likelihood (ML) detector, we analyze the single-user average bit error rate (ABER) bound as the benchmark of the ABER performance for our proposed OBSCMA with CoMP system. We also develop a customized Gaussian approximation with expectation propagation (GAEP) algorithm for multi-user detection and propose efficient algorithm structures for centralized and decentralized detectors. Our proposed OBSCMA with CoMP system leads to stronger performance than the existing solutions. The proposed centralized and decentralized detectors exhibit effective reception and robustness under channel state information uncertainty.
\end{abstract}


%
\IEEEpeerreviewmaketitle

\section{Introduction}
%
%
%
%

The demand for the variety and the number of mobile devices along with the mobile communication applications have increased tremendously. Communications in high-mobility scenarios such as high-speed railways and vehicle-to-everything (V2X) suffer from the well-known time-varying channels with high Doppler spread \cite{8486963}. Traditional orthogonal frequency division multiplexing (OFDM) systems experience significant performance loss due to the severe inter-carrier-interference (ICI) caused by the channel Doppler spread. The recent emergence of orthogonal time frequency space (OTFS) modulation \cite{hadani2017orthogonal} can exploit the degrees of freedom coming from both the delay and Doppler channel domains, resulting in superior performance and more reliable communications compared to OFDM. OTFS can effectively simplify channel estimation \cite{shen2019channel,mishra2021otfs,li2022joint} and symbol detection \cite{raviteja2018interference,yuan2021iterative,xiang2021gaussian,thaj2020low} at wireless receivers by utilizing the property of a quasi-stationary sparse channel in delay-Doppler domain for high-mobility communication scenarios. The performance of uncoded and coded OTFS systems has been analyzed and evaluated in \cite{surabhi2019diversity,raviteja2019effective,singh2022ber} and \cite{li2021performance}, respectively.

Meanwhile, efficient multiple access protocols represent another strong arena of growth. Recently, several works \cite{khammammetti2018otfs,augustine2019interleaved,surabhi2019multiple} investigated multiple mobile user access based on OTFS framework with orthogonal resource allocation. In \cite{khammammetti2018otfs,augustine2019interleaved}, the authors proposed to allocate different time-frequency resources to different mobile users with certain impractical assumptions specially the availability of the elusive ideal bi-orthogonal pulses. The authors in \cite{surabhi2019multiple} allocated different delay-Doppler resources to different mobile users and developed efficient receiver algorithm to tackle the significant co-channel interference (CCI).

To further support massive multiple access and explosive transmission needs, non-orthogonal multiple access (NOMA) has been considered as a promising solution for high spectrum efficiency in traditional overloaded multi-user OFDM \cite{sun2017optimal} and multiple-input multiple-output (MIMO) \cite{ding2015application} systems. Existing NOMA methods are mainly categorized into power-domain and code-domain NOMA \cite{ding2017survey,dai2015non}. In particular, sparse code multiple access (SCMA) \cite{dai2019iterative,yuan2018iterative,wang2019near,liu2019spatial} is a code-domain NOMA, which has attracted tremendous research attention due to its excellent performance and low receiver complexity. The application of NOMA to OTFS can effectively improve spectrum utilization and support massive mobile connectivity \cite{chatterjee2020nonorthogonal,deka2021otfs,ding2019otfs,ge2021otfs}. In OTFS-NOMA, multiple mobile users are allowed to share the same delay-Doppler resources simultaneously, and distinguished by either different power levels \cite{chatterjee2020nonorthogonal} or with the help of sparse codewords \cite{deka2021otfs}. The authors in \cite{ding2019otfs} proposed to pair an OTFS mobile user with a group of OFDM stationary users for the implementation of NOMA. Unlike \cite{ding2019otfs}, a recent work \cite{ge2021otfs} suggested an OTFS-based NOMA (OBNOMA) configuration, where multiple stationary users and multiple mobile users are grouped for non-orthogonal channel sharing via only using OTFS modulation. The proposed OBNOMA framework and developed advanced receiver algorithms in \cite{ge2021otfs} can effectively address the CCI and recover the signal for each user.

In addition, multiple antenna systems \cite{mietzner2009multiple,zhang2020prospective} have gained overwhelming research interest in the last several decades both in academia and industry. The integration of OTFS with MIMO systems can acquire additional diversity, multiplexing, and antenna gains compared to conventional single antenna systems \cite{shan2021low,srivastava2021bayesian,qu2022efficient,shi2021deterministic, liu2020uplink,singh2022low,pandey2021low,li2020new}. In \cite{shan2021low}, the authors proposed a low-complexity detector and low-overhead pilot pattern for OTFS system with a large-scale antenna array at the receiver. A Bayesian learning aided simultaneous row and group sparse channel estimation method was proposed in \cite{srivastava2021bayesian} and an efficient symbol detection scheme was proposed in \cite{qu2022efficient} for MIMO-OTFS systems. The authors in \cite{shi2021deterministic} applied a deterministic pilot pattern and developed an efficient channel estimation scheme for downlink massive MIMO-OTFS systems. An uplink-aided high mobility downlink channel estimation method was proposed in \cite{liu2020uplink} for massive MIMO-OTFS networks. The work in \cite{singh2022low} analyzed and evaluated the performance of linear receivers for MIMO-OTFS systems. The authors in \cite{pandey2021low} proposed a low complexity multi-user precoding and detector for downlink massive MIMO-OTFS systems. A novel path division multiple access (PDMA) was proposed in \cite{li2020new}, where the additional angle-domain can be utilized to differentiate mobile users for massive MIMO-OTFS networks.

However, most of the existing works do not fully utilize the potential performance gain by applying restriction to co-located MIMO designs. To preserve sufficient diversity and exploit favorable propagation property of mobile communications, we shall apply distributed antenna systems with coordinated multi-point (CoMP) transmission and reception. CoMP system was originally proposed to improve average spectral efficiency and alleviate interference between intercells in traditional cellular networks \cite{lee2012coordinated}. Here, we adopt CoMP operation between the neighbouring remote radio heads (RRHs) for better exploit of the diversity. Without loss of generality, we focus on a specific scenario in which multiple mobile users are grouped for SCMA, and served by the RRH in front of them and the RRH behind them simultaneously. 
As the signals received by the RRHs in front of and behind the users experience respectively positive and negative Doppler frequency shifts, we propose an OTFS-based SCMA (OBSCMA) with CoMP system to acquire additional diversity coming from Doppler and spatial domain. Due to the sufficient diversity, our proposed OBSCMA with CoMP system can achieve significant performance improvement over its OFDM-based SCMA (OFDM-SCMA) counterparts \cite{lu2017location} for high-mobility communications. 

Specifically, we propose an OBSCMA with CoMP system to support multiple mobile users connectivity, and exploit the underlying channel diversity for better performance. We also develop efficient practical receivers to mitigate CCI and recover the signal for each user. The contributions of our work are summarized as follows:
\begin{enumerate}
\item We propose an OBSCMA with CoMP configuration, where a group of SCMA mobile users modulated in accordance with OTFS, and served by the RRH in front of them and the RRH behind them simultaneously. The proposed system framework can naturally harvest sufficient diversity coming from both the delay, Doppler and spatial domain, and is amenable to effective receiver algorithms. 

\item Based on the maximum likelihood (ML) detector, we analyze the single-user pairwise error probability (PEP) and characterize its average bit error rate (ABER) bound as the benchmark of the ABER performance for our OBSCMA with CoMP system. From the ABER analysis, the performance improvement of the proposed scheme is evaluated. 

\item We develop a practical customized Gaussian approximation with expectation propagation (GAEP) algorithm for multi-user detection. We also propose efficient algorithm structures for centralized and decentralized GAEP detectors to take advantage of underlying channel diversity from the receptions of the RRHs.

\item We demonstrate that our proposed OBSCMA with CoMP system can achieve better performance than the existing solutions such as co-located RRHs and traditional cellular networks, as well as their OFDM-SCMA counterparts. The proposed decentralized detector converges to the results obtained by the centralized one, and both of them exhibit robustness to the imperfect channel state informations (CSIs).
\end{enumerate}

The remainder of this paper is organized as follows. Section \ref{II_model} introduces the system model of the proposed OBSCMA with CoMP system. Section \ref{III_analysis} analyzes the single-user PEP and captures its ABER bound as the benchmark of the OBSCMA with CoMP system. In Section \ref{IV_receiver}, we develop practical receiver algorithms for centralized and decentralized GAEP detectors, and also discuss about their advantages and disadvantages, respectively. The simulation results and discussions are presented in Section \ref{V_simulation}. Finally, Section \ref{VI_conclusion} concludes our work. 

\textit{Notations:} Throughout this paper, we use boldface upper case letters to denote matrices and boldface lower case letters to denote vectors. The superscripts ${\left(  \cdot  \right)^T}$, ${\left(  \cdot  \right)^H}$ and ${\left(  \cdot  \right)^{ - 1}}$ stand for the transpose, conjugate transpose and inverse, respectively. $\left|  \cdot  \right|$ denotes the absolute value of a complex scalar or the cardinality of a set. $\Pr \left(  \cdot  \right)$ defines probability of an event and $Q\left(  \cdot  \right)$ is the tail distribution function of the standard Gaussian distribution. $\exp \left(  \cdot  \right)$ and $\mathbb{E}\left(  \cdot  \right)$ refer to natural exponential function and statistical expectation operation, respectively. ${\left[  \cdot  \right]_N}$ denotes the modulo $N$ operation and $\left\lceil  \cdot  \right\rceil$ represents the round up operation. The imaginary unit is denoted by $j = \sqrt { - 1}$. A complex Gaussian distribution with mean $\bm {\mu}$ and covariance $\bm{\eta}$ is denoted by ${\cal C}{\cal N}\left( {\bm {\mu} ,\bm{\eta} } \right)$. ${\mathbb{C}^{M \times N}}$, ${{\bf{I}}_M}$ and ${{\bf{0}}_{M \times N}}$ denote the ${M \times N}$ matrix with complex entries, ${M \times M}$ identity matrix and ${M \times N}$ zero matrix, respectively. $A\left[ {i,j} \right]$ is the $\left( {i,j} \right)$-th entry of matrix ${\bf{A}}$ and $a\left[ i \right]$ is the $i$-th element of vector ${\bf{a}}$. $\text{diag}\{ {\bf{a}}\}$ denotes a diagonal matrix whose main diagonal is ${\bf{a}}$. The matrix ${{\bf{F}}_N} = {\left( {\frac{1}{{\sqrt N }}{e^{ - j2\pi \frac{{k\ell}}{N}}}} \right)_{k,\ell = 0,1, \cdots ,N - 1}}$ represents the normalized $N$-point discrete Fourier transform (DFT) matrix. The big-O notation $\mathcal{O}\left(  \cdot  \right)$ asymptotically describes the order of computational complexity.

\section{System Model}\label{II_model}
\begin{figure*}
  \centering
  \includegraphics[width=4.8in]{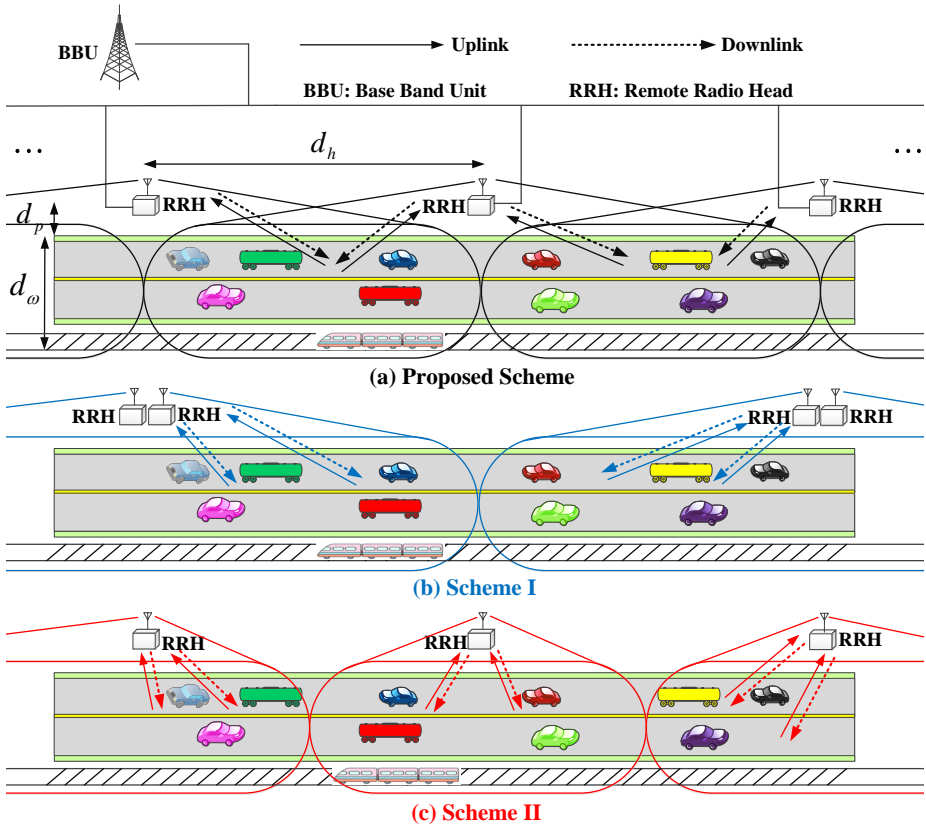}
  \caption{Uplink system model. (a) Proposed OBSCMA with CoMP system. (b) Co-located RRHs Scheme (Scheme I). (c) Traditional cellular network scheme (Scheme II).}
  \label{fig1_CoMP}
\end{figure*}
In this section, we introduce the system model and mathematical expression of the proposed OBSCMA with CoMP system, which includes the system description, transmitter model, channel model and receiver model, respectively.

\subsection{System Description}
We consider an uplink CoMP system shown in Fig. \ref{fig1_CoMP}(a), where the base band unit (BBU) is connected with a serial of RRHs\footnote{The terminology Road Side Unit (RSU) is also commonly found in vehicle communications.} by optical fibers. At
each transmit slot, $J$ independent mobile users located in the same cell, are served by the RRH in front of them and the RRH behind them simultaneously. 
To highlight the superiority of the proposed scheme, we also introduce the co-located RRHs scheme shown in Fig. \ref{fig1_CoMP}(b) and traditional cellular network scheme shown in Fig. \ref{fig1_CoMP}(c) as benchmarks. In particular, the RRHs are co-located deployed and the users are served by the nearest RRHs in Fig. \ref{fig1_CoMP}(b), and the RRHs are distributed deployed and each one only serves the users in its own cell in Fig. \ref{fig1_CoMP}(c). More detailed discussions and comparisons of the different schemes can be found in Section \ref{III_analysis} and Section \ref{V_simulation}.

The message from each user is mapped into a $K$-dimensional SCMA codeword. We assume that each user employs only one SCMA layer and $J>K$ typically, resulting in an overloading factor $\delta  = \frac{J}{K} > 1$. We further assign the SCMA codewords over the delay-Doppler plane and adopt OTFS modulation for uplink transmission. Without loss of generality, a lattice in delay-Doppler plane is denoted as 
$$\Gamma  = \left\{ {\left(\frac{\ell}{{M\Delta f}},\frac{k}{{NT}}\right),\ell = 0, \cdots ,M\! -\! 1;k =\! 0, \!\cdots \!,N \!-\! 1} \right\},$$
and the corresponding time-frequency plane is given by 
$$\Lambda  = \left\{ {(m\Delta f,nT),m = 0, \cdots ,M - 1;n = 0, \cdots ,N - 1} \right\},
$$
where $M$ and $N$ denote, respectively, the total available numbers of subcarriers and time intervals. The choices of $T$ and $\Delta f = {1 \mathord{\left/
 {\vphantom {1 T}} \right.
 \kern-\nulldelimiterspace} T}$ (Hz) should be larger than the maximal channel delay spread and maximum Doppler frequency shift, respectively.
To avoid unnecessary confusion, we use a simple model in which the mobile users and the RRH receivers are equipped with a single transmit antenna and receive antenna. It is worth mentioning that our proposed scheme also applies to the scenarios involving multiple transmit and receive antennas, with expected performance gain. We also emphasize that the performance advantages of the proposed scheme and algorithm still work for downlink scenario in a straightforward manner.

\subsection{Transmitter Model}
At the transmitter, every ${\log _2}Q$ information bits ${{\bf{b}}_j}$ from the $j$-th user are mapped into a complex $K$-dimensional sparse codeword ${{\bf{c}}_j} = {\left[ {{c_{j,1}},{c_{j,2}}, \cdots ,{c_{j,K}}} \right]^T}$ selected from a user-specific SCMA codebook $\mathbb{A}_j$ of size $Q$, where $j = \{ 1,2, \cdots ,J\} $. We assume that only $D<K$ non-zero entries among a $K$-dimensional codeword ${{\bf{c}}_j}$. We then generate the information symbols ${\bf{X}}_j \in {\mathbb{C}^{M \times N}}$ of the $j$-th user by allocating $\frac{{MN}}{K}$ SCMA codewords ${{\bf{c}}_j}$ over the delay-Doppler plane $\Gamma$ without overlapping.

\begin{figure}
  \centering
  \includegraphics[width=3.5in]{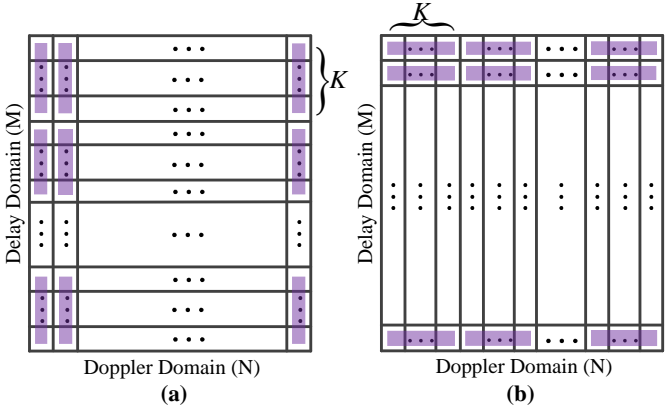}
  \caption{SCMA codewords allocation. (a) Allocated along the delay axis. (b) Allocated along the Doppler axis.}
  \label{fig2_allocation}
\end{figure}

In general, the SCMA codewords can be allocated either along the delay axis in Fig. \ref{fig2_allocation}(a) or along the Doppler axis in Fig. \ref{fig2_allocation}(b). For simplicity, here, we assume that $M$ and $N$ are integer multiples of $K$, i.e., ${[M]_K} = {[N]_K} = 0$, where ${\left[  \cdot  \right]_k}$ denotes mod-$k$ operation.

The delay-Doppler symbols ${\bf{X}}_j \in {\mathbb{C}^{M \times N}}$ are then converted into a lattice in time-frequency domain ${\bf{\bar X}}_j\in {\mathbb{C}^{M \times N}}$ through the inverse symplectic finite Fourier transform (ISFFT) for each user,
\begin{align}\label{ISFFT}
{\bf{\bar X}}_j = {{\bf{F}}_M}{\bf{X_jF}}_N^H,
\end{align}
where ${{\bf{F}}_M} \in {\mathbb{C}^{M \times M}}$ and ${{\bf{F}}_N} \in {\mathbb{C}^{N \times N}}$ denote, respectively, the normalized $M$-point and $N$-point DFT matrices.

Next, each time-frequency signal ${{\bf{\bar X}}_j}$ is transformed into a time domain signal ${\bf{s}}_j \in {\mathbb{C}^{MN \times 1}}$ through Heisenberg transform with a transmit pulse ${g_{tx}}(t)$,
\begin{align}
s_j[c] =& \sum\limits_{n = 0}^{N - 1} {\sum\limits_{m = 0}^{M - 1} {{\bar X}_j[m,n]{g_{tx}}(c{T_s} - nT){e^{j2\pi m \Delta f(c{T_s} - nT)}}} }, \nonumber \\ \;& \quad\quad c = 0, \cdots, MN - 1,
\end{align}
where ${T_s} = {1 \mathord{\left/
 {\vphantom {1 {M\Delta f}}} \right.
 \kern-\nulldelimiterspace} {M\Delta f}}$ is the sampling interval.
 
We then apply a cyclic prefix (CP) in front of the generated time domain signal for each user. After passing through a transmit filter, each mobile user signal is sent out over a doubly-selective fading channel.

\subsection{Channel Model}

In this work, we characterize the channel between $j$-th user and $u$-th RRH as
\begin{align}\label{channel_distant}
{g_{uj}}[c,p] = \sqrt {PL({d_{uj}})} {h_{uj}}[c,p], \;&c = 0, \cdots, MN - 1;\nonumber\\ & p = 0, \cdots ,P_{uj} - 1,
\end{align}
where ${PL({d_{uj}})}$ represents the distance-dependent pathloss, ${{d_{uj}}}$ is the distance from $j$-th user to the $u$-th RRH,\footnote{We assume that the distance remains constant during an OTFS transmission frame.} and $u = \{ 1,2\} $. 

In (\ref{channel_distant}), ${{\bf{h}}_{uj}}$ represents the time-varying multipath fading channel with sampled impulse response
\begin{align}\label{channel_multipath}
{h_{uj}}\left[ {c,p} \right]
= \sum\limits_{i = 1}^{L_{uj}} {{h_{uj,i}}{e^{j2\pi {\nu _{uj,i}}\left( {c{T_s} - p{T_s}} \right)}}{{\mathop{\rm P}\nolimits} _\text{{rc}}}(p{T_s} - {t_{uj}}-{\tau _{uj,i}})}, 
\end{align}
where ${L_{uj}}$ and ${t_{uj}}$ denote the number of multipaths and the amount of timing offset between the $j$-th user and $u$-th RRH; $h_{uj,i}$, $\tau _{uj,i}$ and $\nu _{uj,i}$ are the corresponding channel gain, delay and Doppler frequency shift associated with the $i$-th path, respectively. The Doppler frequency shift of the $i$-th path can be further written as $\nu _{uj,i} = ({{k_{{uj,i}}} + {\beta _{{uj,i}}}})/NT$, where integer ${{k_{{uj,i}}}}$ and real
${\beta _{{uj,i}}} \in \left( { -0.5,0.5} \right]$ denote the index and fractional part of $\nu _{uj,i}$, respectively.

The maximal channel tap $P_{uj}$ is determined by the duration of the filter response and the maximum channel delay spread. In general, the implemented pulse shaping filters at the transmitter and receiver are the root-raised-cosine (RRC) filters, leading to an equivalent overall raised-cosine (RC) rolloff pulse for ${{{\mathop{\rm P}\nolimits} _\text{{rc}}}(\tau )}$ in (\ref{channel_multipath}). In addition, we assume that the CP is long enough to accommodate both the maximum timing offset and the maximal channel delay spread for all users. Hence, there is no inter-frame interference.

\subsection{Receiver Model}

At the receiver, the CP is removed after the time domain signal enters a received filter. We can express the received signal from the $j$-th user at the $u$-th RRH as
\begin{align}
r_{uj}[c] = \sum\limits_{p = 0}^{P_{uj} - 1} {g_{uj}[c,p]s_j\left[ {{{\left[ {c - p} \right]}_{MN}}} \right]},\; c = 0, \cdots ,MN - 1.
\end{align}
The resulting time domain signal ${\bf{r}}_{uj} \in {\mathbb{C}^{MN \times 1}}$ is then transformed into the time-frequency domain by Wigner transform with a receive pulse ${g_{rx}}(t)$,
\begin{align}
\bar Y_{uj}[m,n] = &\sum\limits_{c = 0}^{MN - 1} {g_{rx}^*(c{T_s} - nT)r_{uj}[c]{e^{ - j2\pi m\Delta f(c{T_s} - nT)}}}, \nonumber \\ &{m = 0, \cdots ,M - 1};\; {n = 0, \cdots ,N - 1}.
\end{align}

Finally, the time-frequency signal ${\bf{\bar Y}}_{uj} \in {\mathbb{C}^{M \times N}}$ is transformed back to delay-Doppler domain via the symplectic finite Fourier transform (SFFT),
\begin{align}
{\bf{Y}}_{uj} = {\bf{F}}_M^H{\bf{\bar Y}}_{uj}{{\bf{F}}_N}.
\end{align}

For analytical convenience, we adopt a rectangular pulse for ${g_{tx}}(t)$ and ${g_{rx}}(t)$ in the above steps, and express the baseband OTFS input-output relationship from $j$-th user to $u$-th RRH in delay-Doppler domain as \cite{ge2021receiver}
\begin{align}\label{relationship_DD}
&Y_{uj}[\ell,k] \!=\! \sqrt {PL({d_{uj}})} \sum\limits_{p = 0}^{P_{uj} \!- \!1} {\sum\limits_{i = 1}^{L_{uj}} {\sum\limits_{q = 0}^{N\! -\! 1} {{h_{uj,i}}{{\mathop{\rm P}\nolimits} _\text{{rc}}}(p{T_s}\! -\! {t_{uj}}\!-\!{\tau _{uj,i}})} } } \nonumber\\
&\quad \times \gamma (k,\ell,p,q,{k_{{uj,i}}},{\beta _{{uj,i}}}) X_j\left[ {{{\left[ {\ell - p} \right]}_M},{{\left[ {k - {k_{{uj,i}}} + q} \right]}_N}} \right],
\end{align}
where
\begin{subequations}
\begin{equation}
\begin{aligned}
&\gamma (k,\ell,p,q,{k_{{uj,i}}},{\beta _{{uj,i}}}) \\
&=
\begin{cases}
\frac{1}{N}\xi (\ell,p,{k_{{uj,i}}},{\beta _{{uj,i}}})\theta (q,{\beta _{{uj,i}}}),&p \!\le\! \ell \!<\! M,\\
\frac{1}{N}\xi (\ell,p,{k_{{uj,i}}},{\beta _{{uj,i}}})\theta (q,{\beta _{{uj,i}}})\phi (k,q,{k_{{uj,i}}}), &0 \!\le\! \ell\! <\! p,
\end{cases}
\end{aligned}
\end{equation}
\begin{align}
\xi (\ell,p,{k_{{uj,i}}},{\beta _{{uj,i}}}) = {e^{j2\pi \left( {\frac{{\ell - p}}{M}} \right)\left( {\frac{{{k_{{uj,i}}} + {\beta _{{uj,i}}}}}{N}} \right)}},
\end{align}
\begin{align}
\theta (q,{\beta _{{uj,i}}}) = \frac{{{e^{ - j2\pi ( - q - {\beta _{{uj,i}}})}} - 1}}{{{e^{ - j\frac{{2\pi }}{N}( - q - {\beta _{{uj,i}}})}} - 1}},
\end{align}
\begin{align}
\phi (k,q,{k_{{uj,i}}}) = {e^{ - j2\pi \frac{{{{\left[ {k - {k_{{uj,i}}} + q} \right]}_N}}}{N}}}.
\end{align}
\end{subequations}

The input-output relationship in (\ref{relationship_DD}) can be further column-wise vectorized as
\begin{align}
{{\bf{y}}_{uj}} = {{\bf{H}}_{uj}}{{{\bf{\tilde x}}}_j},
\end{align}
where ${{{\bf{\tilde x}}}_j},{{\bf{y}}_{uj}}\in {\mathbb{C}^{MN \times 1}}$, and ${{\bf{H}}_{uj}}\in {\mathbb{C}^{MN \times MN}}$ is a sparse matrix.

\begin{figure*}
  \centering
  \includegraphics[width=6.8in]{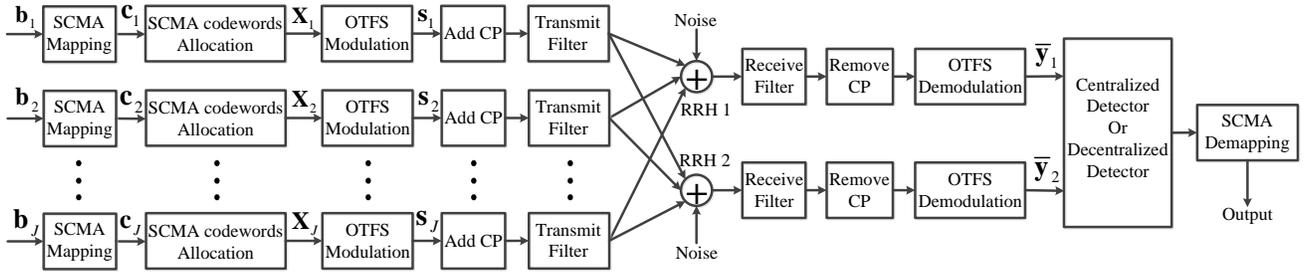}
  \caption{Block diagram of the proposed OBSCMA with CoMP system.}
  \label{fig3_diagram}
\end{figure*}

Consequently, the observations obtained at $u$-th RRH can be expressed as
\begin{align}\label{relation_singlesum}
{{{\bf{\bar y}}}_u} = \sum\limits_{j = 1}^J {\sqrt {{P_j}} {{\bf{H}}_{uj}}{{{\bf{\tilde x}}}_j}}  + {{\bm{\omega }}_u},\;u = 1,2,
\end{align}
where ${{\bm{\omega }}_u}\in {\mathbb{C}^{MN \times 1}}\sim \mathcal{CN}\left( {{\bf{0}},{N_0}{\bf{I}}} \right)$ is the complex additive white Gaussian noise (AWGN) at $u$-th RRH, and ${{P_j}}$ is the transmission power of $j$-th user. The basic block diagram of the considered system is illustrated in Fig. \ref{fig3_diagram}.

\section{Performance Analysis}\label{III_analysis}

As shown by \cite{liu2019spatial,lim2017uplink,bao2017performance} and the results of this section, the single-user ABER performance bound can actually be regarded as the benchmark of the ABER performance for the multi-user scenario with the optimal ML detector. In order to make the analysis tractable, in this section, we analyze the single-user PEP and characterize its ABER bound as the benchmark of our OBSCMA with CoMP system.

When the proposed OBSCMA with CoMP system supports only a single user ($j$-th user is considered here without loss of generality), the received signal at $u$-th RRH in (\ref{relation_singlesum}) can be rewritten in an alternate form as
\begin{align}\label{analys_single}
{{\bf{\bar y}}_u} = \sqrt {{P_j} \times PL\left( {{d_{uj}}} \right)} {{\bf{\Phi }}^{(u)}}\left( {{{\bf{X}}_j}} \right){{\bf{h}}_{uj}} + {{\bm{\omega }}_u},\;u = 1,2,
\end{align}
where ${{\bf{h}}_{uj}} = {\left[ {{h_{uj,1}},{h_{uj,2}} \cdots ,{h_{uj,{L_{uj}}}}} \right]^T}\in {\mathbb{C}^{{L_{uj}} \times 1}}$ is the channel coefficient vector between the $j$-th user and $u$-th RRH, and ${{\bf{\Phi }}^{(u)}}\left( {{{\bf{X}}_j}} \right)\in {\mathbb{C}^{MN \times {L_{uj}}}}$ is referred to as the equivalent transmitted codeword matrix whose $i$-th row ($i = kM + \ell,i = 0,1, \cdots ,MN - 1$) is given in (\ref{transmit_row}), as shown at the top of the next page.
\begin{figure*}[!t]
\begin{align}\label{transmit_row}
\Phi _i^{(u)}\left( {{{\bf{X}}_j}} \right) = \left[
 \begin{matrix}
   {\sum\limits_{p = 0}^{{P_{uj}} - 1} {\sum\limits_{q = 0}^{N - 1} {{{\rm{P}}_\text{rc}}(p{T_s} - {t_{uj}} - {\tau _{uj,1}})\gamma (k,\ell ,p,q,{k_{uj,1}},{\beta _{uj,1}}){X_j}\left[ {{{\left[ {\ell  - p} \right]}_M},{{\left[ {k - {k_{uj,1}} + q} \right]}_N}} \right]} } } \\
   {\sum\limits_{p = 0}^{{P_{uj}} - 1} {\sum\limits_{q = 0}^{N - 1} {{{\rm{P}}_\text{rc}}(p{T_s} - {t_{uj}} - {\tau _{uj,2}})\gamma (k,\ell ,p,q,{k_{uj,2}},{\beta _{uj,2}}){X_j}\left[ {{{\left[ {\ell  - p} \right]}_M},{{\left[ {k - {k_{uj,2}} + q} \right]}_N}} \right]} } } \\
   \vdots \\
   {\sum\limits_{p = 0}^{{P_{uj}} - 1} {\sum\limits_{q = 0}^{N - 1} {{{\rm{P}}_\text{rc}}(p{T_s} - {t_{uj}} - {\tau _{uj,{L_{uj}}}})\gamma (k,\ell ,p,q,{k_{uj,{L_{uj}}}},{\beta _{uj,{L_{uj}}}}){X_j}\left[ {{{\left[ {\ell  - p} \right]}_M},{{\left[ {k - {k_{uj,{L_{uj}}}} + q} \right]}_N}} \right]} } } 
  \end{matrix}
  \right]^T.
\end{align}
	\hrule
\end{figure*}

In order to investigate the theoretical error performance bound, we combine the equations in (\ref{analys_single}) as
\begin{align}
{\bf{\bar y}} = \sqrt {{P_j}} {\bf{\Phi }}\left( {{{\bf{X}}_j}} \right){{\bf{h}}_j} + {\bm{\omega }},
\end{align}
where ${{\bf{\bar y}}} = {\left[ {{\bf{\bar y}}_1^T,{\bf{\bar y}}_2^T} \right]^T}\in {\mathbb{C}^{2MN \times 1}}$, ${{\bf{h}}_j} = {\left[ {{\bf{h}}_{1j}^T,{\bf{h}}_{2j}^T} \right]^T}\in {\mathbb{C}^{\left( {{L_{1j}} + {L_{2j}}} \right) \times 1}}$, $\bm{\omega }={\left[ {{\bm{\omega }}_1^T,{\bm{\omega }}_2^T} \right]^T}\in {\mathbb{C}^{2MN \times 1}}$, and ${\bf{\Phi }}\left( {{{\bf{X}}_j}} \right) = \left[
 \begin{matrix}
   {\sqrt {PL\left( {{d_{1j}}} \right)} {{\bf{\Phi }}^{(1)}}\left( {{{\bf{X}}_j}} \right)} & {\bf{0}}  \\
   {\bf{0}} & {\sqrt {PL\left( {{d_{2j}}} \right)} {{\bf{\Phi }}^{(2)}}\left( {{{\bf{X}}_j}} \right)}
  \end{matrix}
  \right]\in {\mathbb{C}^{2MN \times \left( {{L_{1j}} + {L_{2j}}} \right)}}$.

Assuming perfect CSI is available at the receiver, the conditional PEP, i.e., the probability of transmitting ${{{\bf{X}}_j}}$ but erroneously deciding on ${{{\bf{\hat X}}}_j}$, is given by
\begin{align}\label{analys_cond}
\Pr &\left( {{{\bf{X}}_j} \to {{{\bf{\hat X}}}_j}\left| {{d_{1j}},{d_{2j}},{{\bf{h}}_j}} \right.} \right) \nonumber\\ &= Q\left( {\sqrt {\frac{\rho }{2}{{\left\| {\left( {{\bf{\Phi }}\left( {{{\bf{X}}_j}} \right) - {\bf{\Phi }}\left( {{{{\bf{\hat X}}}_j}} \right)} \right){{\bf{h}}_j}} \right\|}^2}} } \right),
\end{align}
where $Q\left( x \right)$ is the tail distribution function of the standard Gaussian distribution and $\rho  = {{{P_j}} \mathord{\left/
 {\vphantom {{{P_j}} {{N_0}}}} \right.
 \kern-\nulldelimiterspace} {{N_0}}}$.

By resorting the approximation $Q\left( x \right) \approx \frac{1}{{12}}\exp ({{ - {x^2}} \mathord{\left/
 {\vphantom {{ - {x^2}} 2}} \right.
 \kern-\nulldelimiterspace} 2}) + \frac{1}{4}\exp ({{ - 2{x^2}} \mathord{\left/
 {\vphantom {{ - 2{x^2}} 3}} \right.
 \kern-\nulldelimiterspace} 3})$, we can rewrite (\ref{analys_cond}) as
\begin{align}\label{analys_cond_appro}
\Pr& \left( {{{\bf{X}}_j} \to {{{\bf{\hat X}}}_j}\left| {{d_{1j}},{d_{2j}},{{\bf{h}}_j}} \right.} \right)\nonumber\\ &\approx \frac{1}{{12}}\exp \left( { - \frac{{\rho {{\left\| {\left( {{\bf{\Phi }}\left( {{{\bf{X}}_j}} \right) - {\bf{\Phi }}\left( {{{{\bf{\hat X}}}_j}} \right)} \right){{\bf{h}}_j}} \right\|}^2}}}{4}} \right) \nonumber\\
&\quad + \frac{1}{4}\exp \left( { - \frac{{\rho {{\left\| {\left( {{\bf{\Phi }}\left( {{{\bf{X}}_j}} \right) - {\bf{\Phi }}\left( {{{{\bf{\hat X}}}_j}} \right)} \right){{\bf{h}}_j}} \right\|}^2}}}{3}} \right).
\end{align}

After averaging over the distances ${{d_{1j}}}$ and ${{d_{2j}}}$, (\ref{analys_cond_appro}) can be approximately expressed as 
\begin{align}\label{analys_cond_final}
\Pr& \left( {{{\bf{X}}_j} \to {{{\bf{\hat X}}}_j}\left| {{{\bf{h}}_j}} \right.} \right) \nonumber\\ &\approx  \frac{1}{{12}}\exp \left( { - \frac{{\rho {{\left\| {\left( {{\bf{\bar \Phi }}\left( {{{\bf{X}}_j}} \right) - {\bf{\bar \Phi }}\left( {{{{\bf{\hat X}}}_j}} \right)} \right){{\bf{h}}_j}} \right\|}^2}}}{4}} \right) \nonumber\\
& \quad + \frac{1}{4}\exp \left( { - \frac{{\rho {{\left\| {\left( {{\bf{\bar \Phi }}\left( {{{\bf{X}}_j}} \right) - {\bf{\bar \Phi }}\left( {{{{\bf{\hat X}}}_j}} \right)} \right){{\bf{h}}_j}} \right\|}^2}}}{3}} \right),
\end{align}
where\\
${{\bf{\bar \Phi }}\left( {{{\bf{X}}_j}} \right)} \!=\! \left[
 \begin{matrix}
   {\sqrt {\mathbb{E}\left( {PL\left( {{d_{1j}}} \right)} \right)} {{\bf{\Phi }}^{(1)}}\left( {{{\bf{X}}_j}} \right)}\! &\!\! {\bf{0}}  \\
   {\bf{0}}\! & \!\!{\sqrt {\mathbb{E}\left( {PL\left( {{d_{2j}}} \right)} \right)} {{\bf{\Phi }}^{(2)}}\left( {{{\bf{X}}_j}} \right)}
  \end{matrix}
  \right]$, and $\mathbb{E}( \cdot )$ represents expectation.

Note that ${\left( {{\bf{\bar \Phi }}\left( {{{\bf{X}}_j}} \right) - {\bf{\bar \Phi }}\left( {{{{\bf{\hat X}}}_j}} \right)} \right)^H}\left( {{\bf{\bar \Phi }}\left( {{{\bf{X}}_j}} \right) - {\bf{\bar \Phi }}\left( {{{{\bf{\hat X}}}_j}} \right)} \right)$ is a Hermitian matrix, its rank and the non-zero eigenvalues are defined as $R$ and ${\lambda _i},i = 1,2, \cdots ,R$, respectively. Hence, we can obtain 
\begin{align}
&{\left\| {\left( {{\bf{\bar \Phi }}\left( {{{\bf{X}}_j}} \right) - {\bf{\bar \Phi }}\left( {{{{\bf{\hat X}}}_j}} \right)} \right){{\bf{h}}_j}} \right\|^2} \nonumber\\ =& {\bf{h}}_j^H{\left( {{\bf{\bar \Phi }}\left( {{{\bf{X}}_j}} \right) - {\bf{\bar \Phi }}\left( {{{{\bf{\hat X}}}_j}} \right)} \right)^H}\left( {{\bf{\bar \Phi }}\left( {{{\bf{X}}_j}} \right) - {\bf{\bar \Phi }}\left( {{{{\bf{\hat X}}}_j}} \right)} \right){{\bf{h}}_j}\nonumber\\
 =& {\bf{h}}_j^H{\bf{U\Sigma }}{{\bf{U}}^H}{{\bf{h}}_j}\nonumber\\
 =& {\bf{\tilde h}}_j^H{\bf{\Sigma }}{{{\bf{\tilde h}}}_j}\nonumber\\
 =& \sum\limits_{i = 1}^R {{\lambda _i}{{\left| {{{\tilde h}_{ji}}} \right|}^2}},
\end{align}
where ${\bf{U}}$ is a unitary matrix, ${{{\bf{\tilde h}}}_j} = {\bf{h}}_j^H{\bf{U}}$ and ${\bf{\Sigma }} = \text{diag}\{ {\lambda _1},{\lambda _2}, \cdots ,{\lambda _{{L_{1j}} + {L_{2j}}}}\}$.

Since ${{{\bf{\tilde h}}}_j}$ is obtained by multiplying a unitary matrix with ${\bf{h}}_j^H$, it has the same distribution as that of ${\bf{h}}_j^H$. The elements in ${{{\bf{\tilde h}}}_j}$ are assumed to be independent and identically distributed complex Gaussian random variables. Considering ${{{\bf{\tilde h}}}_j} \sim \mathcal{CN}\left( {{\bf{0}},\frac{2}{{{L_{1j}} + {L_{2j}}}}{{\bf{I}}_{{L_{1j}} + {L_{2j}}}}} \right)$, the final PEP is calculated by averaging (\ref{analys_cond_final}) over the channel statistics and given by
\begin{align}\label{analys_uncond}
\Pr& \left( {{{\bf{X}}_j} \to {{{\bf{\hat X}}}_j}} \right) \nonumber\\& \approx \frac{1}{{12}}\mathop \prod \limits_{i = 1}^R \frac{1}{{1 + \frac{{\rho {\lambda _i}}}{{2\left( {{L_{1j}} + {L_{2j}}} \right)}}}} + \frac{1}{4}\mathop \prod \limits_{i = 1}^R \frac{1}{{1 + \frac{{2\rho {\lambda _i}}}{{3\left( {{L_{1j}} + {L_{2j}}} \right)}}}}.
\end{align}
As $\rho  \to \infty $, (\ref{analys_uncond}) can be further simplified as
\begin{align}
&\Pr \left( {{{\bf{X}}_j} \to {{{\bf{\hat X}}}_j}} \right) \nonumber\\ &\approx \left[ {\frac{1}{{12}}{{\left( {\frac{{{{\left( {\mathop \prod \limits_{i = 1}^R {\lambda _i}} \right)}^{\frac{1}{R}}}}}{{2\left( {{L_{1j}} \!+\! {L_{2j}}} \right)}}} \right)}^{ - R}} \!+\! \frac{1}{4}{{\left( {\frac{{2{{\left( {\mathop \prod \limits_{i = 1}^R {\lambda _i}} \right)}^{\frac{1}{R}}}}}{{3\left( {{L_{1j}} \!+\! {L_{2j}}} \right)}}} \right)}^{ - R}}} \right]{\rho ^{ - R}}.
\end{align}

From the above analysis, we conclude that the system diversity order is determined by $R$, which could be as high as the summation of the resolvable paths of the channels between the user and the RRHs. Consequently, according to the union bounding technique, the ABER of the single-user system can be upper bounded by 
\begin{align}
&{P_e}\le \nonumber \\ & \frac{1}{{J{Q^{\frac{{MN}}{K}}}\frac{{MN}}{K}{{\log }_2}Q}}\sum\limits_{j \!=\! 1}^J {\sum\limits_{{{\bf{X}}_j}} {\sum\limits_{{{{\bf{\hat X}}}_j} \ne {{\bf{X}}_j}} {\Pr \left( {{{\bf{X}}_j} \!\to\! {{{\bf{\hat X}}}_j}} \right)e\left( {{{\bf{X}}_j},{{{\bf{\hat X}}}_j}} \right)} } },
\end{align}
where ${e\left( {{{\bf{X}}_j},{{{\bf{\hat X}}}_j}} \right)}$ represents the number of bits in difference for the corresponding pairwise error event.

\begin{figure}
  \centering
  \includegraphics[width=3.6in]{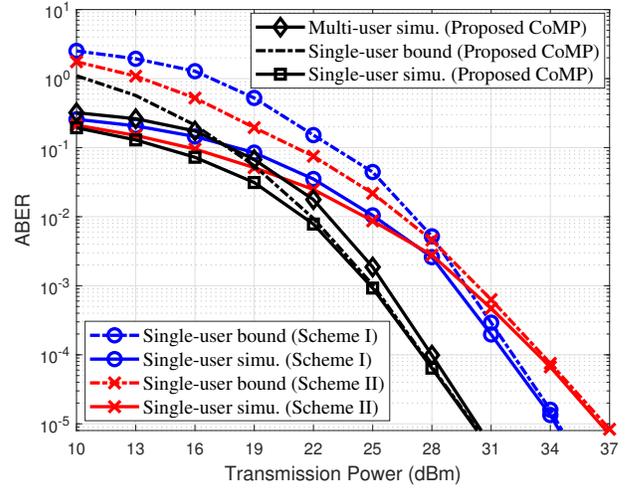}
  \caption{ABER performance comparison for different scenarios.}\label{PEP1}
\end{figure}
In Fig. \ref{PEP1}, we examine the effectiveness of the ABER performance analysis with ML detector. The user-specific codebooks are generated according to \cite{xiao2018capacity} with $J=6$, $K=4$, $D=2$ and $Q=4$. We consider a delay-Doppler plane with $M=4$ and $N=2$. The transmission power of each mobile user is assumed to be the same. Here, we select a typical urban channel model \cite{failli1989digital} with $L_{1j}=L_{2j}=L=4$ resolvable paths for each user. The channel response for each mobile user is generated by using Jakes formulation \cite{ge2021receiver,raviteja2018interference} with maximum Doppler frequency shift equals to 1111 Hz for the speed of 300 km/h. All the other simulation setups are the same as introduced in Section \ref{V_simulation}. To highlight the superiority of the proposed scheme, we also provide the single-user benchmark performance of co-located RRHs scheme shown in Fig. \ref{fig1_CoMP}(b) and traditional cellular network scheme shown in Fig. \ref{fig1_CoMP}(c), denoted as Scheme I and Scheme II, respectively.

From Fig. \ref{PEP1}, we observe that the performance of single-user bound and single-user simulation tend to converge at high region of transmission power for each scheme. Compared to the single-user ABER bound of OBSCMA with CoMP system, the performance of the proposed multi-user simulation degrades in the low transmission power due to the CCI, and achieves a similar ABER result for the high transmission power. Hence, the ABER bound analysis are effective for single-user system and can be applied as the benchmark performance of our OBSCMA with CoMP system. We also notice that the ABER performance of Scheme I and Scheme II intersect with each other as the transmission power increases. The reason is because Scheme II achieves better performance in the low transmission power by leveraging the favorable propagation gain. However, the performance of Scheme I becomes better than Scheme II in the high transmission power due to its benefit of high spatial diversity gain.

\begin{figure}
  \centering
  \includegraphics[width=3.6in]{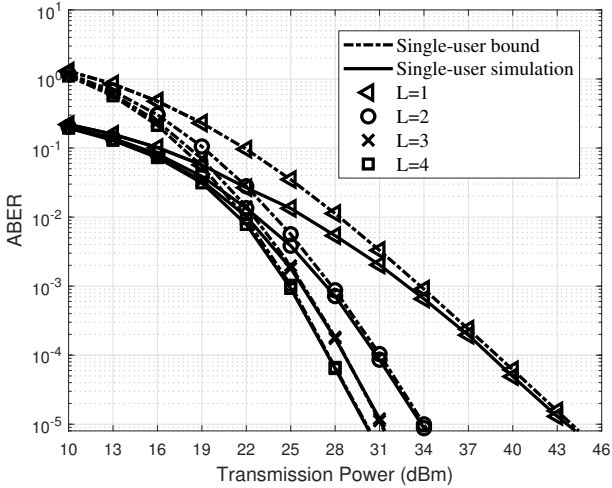}
  \caption{ABER performance comparison with different resolvable paths.}\label{PEP2}
\end{figure}
Fig. \ref{PEP2} further illustrates the effect of the number of resolvable paths on the single-user ABER performance of OBSCMA with CoMP system. It is obvious that the ABER performance improves as $L$ increases. This is due to the fact that high diversity gain can be obtained for better performance with large value of $L$. We again notice that the ABER performance of single-user bound is tight and approach that of single-user simulation for different $L$, which further exhibits the effectiveness of our ABER performance analysis.

\section{Receiver Design}\label{IV_receiver}
In the last section, we characterize the asymptotic single-user ABER performance as the benchmark of our proposed OBSCMA with CoMP system by PEP analysis based on ML detector. However, the complexity order of the ML is $\mathcal{O}\left( {{Q^{\frac{{MNJ}}{K}}}} \right)$, resulting in intolerable computational burden to the receiver for practical large dimensional systems with massive number of user connections. We now investigate practical receiver algorithms to recover the signal for each user from the receptions of the RRHs. To this end, we rewrite the expression in (\ref{relation_singlesum}) as
\begin{align}\label{relation_single}
{{{\bf{\bar y}}}_u} = {{\bf{\bar H}}_u}{{\bf{\bar x}}} + {\bm{\omega }_u},\;u = 1,2,
\end{align}
where ${{{\bf{\bar H}}}_u} = \left[ {\sqrt {{P_1}} {{\bf{H}}_{u1}},\sqrt {{P_2}} {{\bf{H}}_{u2}}, \cdots ,\sqrt {{P_J}} {{\bf{H}}_{uJ}}} \right]\in {\mathbb{C}^{MN \times MNJ}}$, and ${\bf{\bar x}} = {\left[ {{\bf{\tilde x}}_1^T,{\bf{\tilde x}}_2^T, \cdots ,{\bf{\tilde x}}_J^T} \right]^T}\in {\mathbb{C}^{MNJ \times 1}}$.

Note that ${{\bf{\bar x}}}$ is a sparse vector due to the sparse SCMA codewords. The number of non-zero entries in ${{\bf{\bar x}}}$ is only $\frac{{MNJD}}{K}$. Let ${{\bf{\hat x}}}\in {\mathbb{C}^{\frac{{MNJD}}{K} \times 1}}$ denotes the effective input after removing the zeros in ${{\bf{\bar x}}}$, and ${{{\bf{\hat H}}}_u}\in {\mathbb{C}^{MN \times \frac{{MNJD}}{K}}}$ represents the effective matrix after deleting the columns corresponding to the indices of zeros in ${{\bf{\bar x}}}$. Thus, we can simplify the relationship of (\ref{relation_single}) to 
\begin{align}\label{relation_singlesimple}
{{{\bf{\bar y}}}_u} = {{{\bf{\hat H}}}_u}{{\bf{\hat x}}} + {\bm{\omega }_u},\;u = 1,2.
\end{align}

As ${{\bf{\hat x}}}$ contains information from $\frac{{MNJ}}{K}$ SCMA codewords, we further group every $D$ non-zero elements from the same SCMA codeword in ${{\bf{\hat x}}}$. Similarly, the corresponding columns in ${{{\bf{\hat H}}}_u}$ should be grouped together. We can now rewrite (\ref{relation_singlesimple}) as
\begin{align}\label{relation_singlefinal}
{{{\bf{\bar y}}}_u} = {{{\bf{H}}}_u}{{\bf{x}}} + {\bm{\omega }_u},\;u = 1,2,
\end{align}
where ${\bf{x}} = {\left[ {{\bf{x}}_1^T,{\bf{x}}_2^T, \cdots ,{\bf{x}}_{{{MNJ} \mathord{\left/
 {\vphantom {{MNJ} K}} \right.
 \kern-\nulldelimiterspace} K}}^T} \right]^T}\in {\mathbb{C}^{\frac{{MNJD}}{K} \times 1}}$, ${{\bf{x}}_c}\in {\mathbb{C}^{D \times 1}}$, ${\bf{h}}_{d,c}^u\in {\mathbb{C}^{1 \times D}}$ and $${{\bf{H}}_u} \!= \!\!\left[ {\begin{array}{*{20}{c}}
\!\!{{\bf{h}}_{1,1}^u}&\!{{\bf{h}}_{1,2}^u}& \cdots &{{\bf{h}}_{1,{{MNJ} \mathord{\left/
 {\vphantom {{MNJ} K}} \right.
 \kern-\nulldelimiterspace} K}}^u}\!\!\\
\!\!{{\bf{h}}_{2,1}^u}&\!{{\bf{h}}_{2,2}^u}& \cdots &{{\bf{h}}_{2,{{MNJ} \mathord{\left/
 {\vphantom {{MNJ} K}} \right.
 \kern-\nulldelimiterspace} K}}^u}\!\!\\
\!\! \vdots & \!\vdots & \ddots & \vdots \!\!\\
\!\!{{\bf{h}}_{MN,1}^u}&\!{{\bf{h}}_{MN,2}^u}& \cdots &{{\bf{h}}_{MN,{{MNJ} \mathord{\left/
 {\vphantom {{MNJ} K}} \right.
 \kern-\nulldelimiterspace} K}}^u}\!\!
\end{array}} \right]\!\!\!\in\! {\mathbb{C}^{MN \times \frac{{MNJD}}{K}}}.$$

From (\ref{relation_singlefinal}), we observe that the dimension of receptions at each RRH is less than the number of transmitted SCMA codewords as $J>K$. Hence, the conventional multi-user detection for orthogonal multiple access \cite{khammammetti2018otfs,augustine2019interleaved,surabhi2019multiple} cannot be directly applied in such an over-loaded system. To achieve better performance, the advanced receiver algorithms are required to recover the signal of each user. In the sequel, we develop the efficient centralized and decentralized detectors for multi-user detection, and also discuss their advantages and disadvantages, respectively.

\subsection{Centralized Detector}

In this subsection, we introduce the centralized detector to take advantage of the receptions from both RRHs. Specifically, the RRHs will forward their receptions to the BBU for centralized multi-user detection. We can combine the equations in (\ref{relation_singlefinal}) at the BBU as
\begin{align}\label{relation_central}
{\bf{y}} = {\bf{Hx}} + \bm{\omega },
\end{align}
where ${\bf{y}} = {\left[ {{\bf{\bar y}}_1^T,{\bf{\bar y}}_2^T} \right]^T}\in {\mathbb{C}^{2MN \times 1}}$, ${\bf{H}} = {\left[ {{\bf{H}}_1^T,{\bf{H}}_2^T} \right]^T}\in {\mathbb{C}^{2MN \times \frac{{MNJD}}{K}}}$, and $\bm{\omega }={\left[ {{\bm{\omega }}_1^T,{\bm{\omega }}_2^T} \right]^T}\in {\mathbb{C}^{2MN \times 1}}$.

Note that direct solution of (\ref{relation_central}) could be computationally complexity as it involves a large matrix inverse while $MN$ can typically be in the order of thousands or even larger in OTFS system. Fortunately, we can use a sparsely connected factor graph to describe the linear model of (\ref{relation_central}) since $\bf{H}$ is a sparse matrix. The corresponding factor graph includes $2MN$ observation nodes ${y_d}$, $d = 1,2, \cdots ,2MN$, and $\frac{{MNJ}}{K}$ variable nodes ${{\bf{x}}_c}$, $c = 1,2, \cdots ,\frac{{MNJ}}{K}$. An edge is connected between an observation node ${y_d}$ and a variable node ${{\bf{x}}_c}$ if ${{\bf{h}}_{d,c}} \ne {{\bf{0}}_{1 \times D}}$. Let $\mathcal{I}(d)$ and $\mathcal{J}(c)$ denote the index sets of non-zero components (i.e., ${{\bf{h}}_{d,c}} \ne {{\bf{0}}_{1 \times D}}$) in the $d$-th row and $c$-th column of $\bf{H}$, where $d = 1,2, \cdots ,2MN$ and $c = 1,2, \cdots ,\frac{{MNJ}}{K}$, respectively. We also represent the corresponding numbers of non-zero components in the $d$-th row and $c$-th column as $\left| {\mathcal{I}(d)} \right|$ and $\left| {\mathcal{J}(c)} \right|$.

In general, one can derive an efficient MP algorithm \cite{mishra2021otfs,raviteja2018interference,dai2019iterative} for symbol detection according to the sparse factor graph with low complexity. Here, we develop a customized GAEP detector for performance improvement. Recently, EP algorithm is widely applied to approximate the true posterior with modest complexity for traditional MIMO \cite{zhang2020decentralized} and single carrier \cite{csahin2018iterative} systems. The performance advantage of EP motivates us to develop more efficient receiver algorithms in our model to overcome the co-channel user interference and the self-interference in the delay-Doppler domain.
A detailed implementation of the centralized GAEP detector is summarized in \textbf{Algorithm \ref{alg:A}}. We now describe its steps in iteration $\kappa $:
\begin{algorithm}
\caption{Centralized GAEP Detector}
\label{alg:A}
\begin{algorithmic}
\STATE {Input: ${{{\bf{\bar y}}}_1}$, ${{{\bf{\bar y}}}_2}$, ${{\bf{H}}_1}$, ${{\bf{H}}_2}$, ${P_D}\left( {\bf{x}} \right)$ and ${n_c}$.}
\STATE {Initialization: ${{\bm{\mu }}_{d,c}^{(0)}}$, ${\bm{\eta }}_{d,c}^{(0)}$, $c = 1,2, \cdots ,\frac{{MNJ}}{K},d \in \mathcal{J}(c)$, $\delta _I^{(0)}=0$ and iteration count $\kappa=1$.}
\REPEAT
\STATE {1)\; Each observation node ${y_d}$ calculates the mean $C_{d,c}^{(\kappa )}[i]$ and variance $D_{d,c}^{(\kappa )}[i]$ in (\ref{GAEP_y_m}) and (\ref{GAEP_y_v}), and sends them to the connected variable nodes ${{\bf{x}}_c},c \in \mathcal{I}(d)$ if ${{h_{d,c}}[i] \ne 0},i = 1,2, \cdots ,D$;}
\STATE {2)\; Each variable node ${{\bf{x}}_c}$ generates the mean $\mu _{d,c}^{(\kappa )}[i]$ and variance $\eta _{d,c}^{(\kappa )}[i]$ in (\ref{GAEP_x}), and passes them back to the connected observation nodes ${y_d},d \in \mathcal{J}(c)$ if ${{h_{d,c}}[i] \ne 0},i = 1,2, \cdots ,D$;}
\STATE {3)\; Compute the convergence indicator $\delta _I^{(\kappa )}$ in (\ref{conv_c});}
\STATE {4)\; Update $P({\bf{x}}) = {{\bar P}^{(\kappa )}}\left( {\bf{x}} \right)$ if $\delta _I^{(\kappa )} > \delta _I^{(\kappa  - 1)}$;}
\STATE {5)\; $\kappa: = \kappa + 1$;}
\UNTIL{$\delta _I^{(\kappa )}=1$ or $\kappa={n_c}$.}
\STATE {Output: $P({\bf{x}})$.}
\end{algorithmic}
\end{algorithm}

1) From observation node ${y_d}$ to variable nodes ${{\bf{x}}_c},c \in \mathcal{I}(d)$: At each observation node, we can express the received signal ${y_d}$ as
\begin{align}
{y_d} = \sum\limits_{i = 1,{h_{d,c}}[i] \ne 0}^D {{h_{d,c}}[i]{x_c}[i]}  + \sum\limits_{e \in \mathcal{I}(d),e \ne c} {{{\bf{h}}_{d,e}}{{\bf{x}}_e}}  + {\omega _d}.
\end{align}
We approximate the messages updated and passed between the observation nodes and variable nodes on the factor graph as Gaussian. Hence, the observation node ${y_d}$ sends the mean 
$C_{d,c}^{(\kappa )}[i]$ and variance $D_{d,c}^{(\kappa )}[i]$ to the variable node ${{\bf{x}}_c}$ if ${{h_{d,c}}[i] \ne 0},i = 1,2, \cdots ,D$, where
\begin{align}\label{GAEP_y_m}
C_{d,c}^{(\kappa )}[i] = \frac{1}{{{h_{d,c}}[i]}}&\left[ {y_d} - \sum\limits_{e \in \mathcal{I}(d),e \ne c} {{{\bf{h}}_{d,e}}{\bm{\mu }}_{d,e}^{(\kappa  - 1)}} \right. \nonumber\\
&\left. - \sum\limits_{f = 1,f \ne i\hfill\atop
\scriptstyle{h_{d,c}}[f] \ne 0\hfill}^D {{h_{d,c}}[f]\mu _{d,c}^{(\kappa  - 1)}[f]}  \right],
\end{align}
\begin{align}\label{GAEP_y_v}
D_{d,c}^{(\kappa )}[i] = &\frac{1}{{{{\left| {{h_{d,c}}[i]} \right|}^2}}}\left[ \sum\limits_{e \in \mathcal{I}(d),e \ne c} {\sum\limits_{f = 1}^D {{{\left| {{h_{d,e}}[f]} \right|}^2}\eta _{d,e}^{(\kappa  - 1)}[f]} } \right.   \nonumber\\
&\quad \left. + \sum\limits_{f = 1,f \ne i\hfill\atop
\scriptstyle{h_{d,c}}[f] \ne 0\hfill}^D {{{\left| {{h_{d,c}}[f]} \right|}^2}\eta _{d,c}^{(\kappa  - 1)}[f]}  + {\sigma ^2} \right].
\end{align}
Here, ${{\bm{\mu }}_{d,e}^{(\kappa  - 1)}}$ and ${\bm{\eta }}_{d,e}^{(\kappa  - 1)}$ are the mean and variance vectors received from variable node ${{{\bf{x}}_e}}$ in the ${(\kappa  - 1)}$-th iteration. They can be initialized in the first iteration by projecting the equiprobable symbols into Gaussian distribution as (\ref{gaussin_appro}). ${{\sigma ^2}}$ is the variance of the noise at the receiver input.

2) From variable node ${{\bf{x}}_c}$ to observation nodes ${y_d},d \in \mathcal{J}(c)$: The {\em a posteriori} probability can be expressed as follows at each variable node
\begin{align}
&{{\bar P}^{(\kappa )}}\left( {{{\bf{x}}_c} = {{\bm{\chi }}_j}} \right) \nonumber \\ &\!\propto \!{P_D}\left( {{{\bf{x}}_c} \!=\! {{\bm{\chi }}_j}} \right)\!\!\prod\limits_{e \in {\cal J}(c)} {\prod\limits_{i = 1,{h_{e,c}}[i] \ne 0}^D {\exp \left( { \!- \frac{{{{\left| {{\chi _j}[i] \!-\! C_{e,c}^{(\kappa )}[i]} \right|}^2}}}{{D_{e,c}^{(\kappa )}[i]}}} \right)} },\nonumber\\ &\quad\quad \forall {{\bm{\chi }}_j} \in {{\bar {\mathbb{A}}}_j},
\end{align}
where $j = \left\lceil {\frac{{cK}}{{MN}}} \right\rceil $ and $\left\lceil  \cdot \right\rceil $ denotes the round up operation. ${{\bar {\mathbb{A}}}_j}$ is a set contains the non-zero elements of the predefined $j$-th user SCMA codebook $\mathbb{A}_j$, and ${{\bm{\chi }}_j}$ is a $D$-dimensional codeword from ${{\bar {\mathbb{A}}}_j}$. ${P_D}\left( {{{\bf{x}}_c} = {{\bm{\chi }}_j}} \right)$ represents the {\em a priori} probability when ${{{\bf{x}}_c} = {{\bm{\chi }}_j}}$, which can be assumed with equiprobable symbols if no priori information is observed. We then project this posteriori probability into a Gaussian distribution ${\cal C}{\cal N}\left( {E_c^{(\kappa )}[i],F_c^{(\kappa )}[i]} \right), i = 1,2, \cdots ,D$ and set a minimum allowed variance $\varepsilon $, i.e., $F_c^{(\kappa )}[i] = \max \left\{ {\varepsilon ,F_c^{(\kappa )}[i]} \right\}$ to avoid numerical instabilities. The mean ${E_c^{(\kappa )}[i]}$ and variance ${F_c^{(\kappa )}[i]}$ are given by
\begin{subequations}\label{gaussin_appro}
\begin{align}
E_c^{(\kappa )}[i] = \sum\limits_{{{\bm{\chi }}_j} \in {{\bar {\mathbb{A}}}_j}} {{{\bar P}^{(\kappa )}}\left( {{{\bf{x}}_c} = {{\bm{\chi }}_j}} \right){\chi _j}[i]},
\end{align}
\begin{align}
F_c^{(\kappa )}[i] = \sum\limits_{{{\bm{\chi }}_j} \in {{\bar {\mathbb{A}}}_j}} {{{\bar P}^{(\kappa )}}\left( {{{\bf{x}}_c} = {{\bm{\chi }}_j}} \right){{\left| {{\chi _j}[i]} \right|}^2} - {{\left| {E_c^{(\kappa )}[i]} \right|}^2}}.
\end{align}
\end{subequations}
Following the Gaussian message combining rule \cite{ge2021otfs,csahin2018iterative}, we can update the extrinsic distribution 
$\bar q_E^{(\kappa )}\left( {{x_{d,c}}[i]} \right) \sim \mathcal{CN}\left( {\bar \mu _{d,c}^{(\kappa )}[i],\bar \eta _{d,c}^{(\kappa )}[i]} \right)$ if ${{h_{d,c}}[i] \ne 0},i = 1,2, \cdots ,D$, where
\begin{subequations}
\begin{align}
\bar \eta _{d,c}^{(\kappa )}[i] = {\left[ {{{\left( {F_c^{(\kappa )}[i]} \right)}^{ - 1}} - {{\left( {D_{d,c}^{(\kappa )}[i]} \right)}^{ - 1}}} \right]^{ - 1}},
\end{align}
\begin{align}
\bar \mu _{d,c}^{(\kappa )}[i] = \bar \eta _{d,c}^{(\kappa )}[i]\left[ {\frac{{E_c^{(\kappa )}[i]}}{{F_c^{(\kappa )}[i]}} - \frac{{C_{d,c}^{(\kappa )}[i]}}{{D_{d,c}^{(\kappa )}[i]}}} \right].
\end{align}
\end{subequations}
Finally, the variable node ${{\bf{x}}_c}$ calculates the mean $\mu _{d,c}^{(\kappa )}[i]$ and variance $\eta _{d,c}^{(\kappa )}[i]$ as follows and passes them back to the observation node ${y_d}$,
\begin{subequations}\label{GAEP_x}
\begin{align}
\eta _{d,c}^{(\kappa )}[i] = {\left[ {\frac{\Delta }{{\bar \eta _{d,c}^{(\kappa )}[i]}} + \frac{{(1 - \Delta )}}{{\eta _{d,c}^{(\kappa  - 1)}[i]}}} \right]^{ - 1}},
\end{align}
\begin{align}
\mu _{d,c}^{(\kappa )}[i] = \eta _{d,c}^{(\kappa )}[i]\left[ {\Delta \frac{{\bar \mu _{d,c}^{(\kappa )}[i]}}{{\bar \eta _{d,c}^{(\kappa )}[i]}} + (1 - \Delta )\frac{{\mu _{d,c}^{(\kappa  - 1)}[i]}}{{\eta _{d,c}^{(\kappa  - 1)}[i]}}} \right],
\end{align}
\end{subequations}
where $\Delta  \in (0,1]$ is a message damping factor adopted to improve the performance and convergence. If the renewed variance $\eta _{d,c}^{(\kappa )}[i]$ is negative, we ignore the current update and utilize the value of previous iteration instead. 

3) Convergence indicator: The convergence indicator $\delta _I^{(\kappa )}$ is defined as 
\begin{align}\label{conv_c}
\delta _I^{(\kappa )} = &\frac{K}{{MNJ}}\sum\limits_{c = 1}^{{{MNJ} \mathord{\left/
 {\vphantom {{MNJ} K}} \right.
 \kern-\nulldelimiterspace} K}} {\mathbb{I}\left( {\mathop {\max }\limits_{{{\bm{\chi }}_j} \in {{\bar {\mathbb{A}}}_j}} {{\bar P}^{(\kappa )}}\left( {{{\bf{x}}_c} = {{\bm{\chi }}_j}} \right) \ge 1 - \varrho } \right)} ,\nonumber\\ &j = \left\lceil {\frac{{cK}}{{MN}}} \right\rceil
\end{align}
for some small $\varrho>0$ and $\mathbb{I}(\cdot)$ stands for the indicator function.

4) Update criterion: If $\delta _I^{(\kappa )} > \delta _I^{(\kappa  - 1)}$, we then update 
\begin{align}
P({{\bf{x}}_c}) = {{\bar P}^{(\kappa )}}\left( {{{\bf{x}}_c}} \right),\;c = 1,2, \cdots ,\frac{{MNJ}}{K}.
\end{align}

5) Terminating criterion: The centralized GAEP detector terminates if $\delta _I^{(\kappa )}=1$ or the maximum iteration number ${n_c}$ is reached.

We obtain the following decisions once the terminating criterion is satisfied
\begin{align}
{{{\bf{\hat x}}}_c} = \mathop {\arg \max }\limits_{{{\bm{\chi }}_j} \in {{\bar {\mathbb{A}}}_j}} P({{\bf{x}}_c} \!=\! {{\bm{\chi }}_j}),\;j \!=\! \left\lceil {\frac{{cK}}{{MN}}} \right\rceil ,c \!=\! 1,2, \!\cdots\! ,\frac{{MNJ}}{K}.
\end{align}
Finally, the SCMA demapping is applied to recover the transmitted information bits of each user.

\subsection{Decentralized Detector}
\begin{figure}
  \centering
  \includegraphics[width=2.8in]{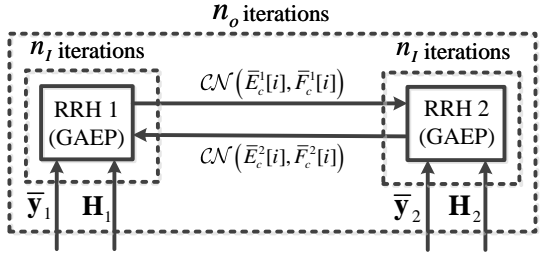}
  \caption{Structure of the decentralized GAEP detector.}
  \label{Fig4_decestructure}
\end{figure}
As the two RRHs are also connected by the optical fibers directly, we can achieve a decentralized processing in a straightforward manner to enable the cooperation between these two RRHs. The structure of the decentralized detector is shown in Fig. \ref{Fig4_decestructure}. Specifically, the two RRHs apply the GAEP algorithm similar to \textbf{Algorithm \ref{alg:A}} for symbol detection independently, and also exchange information iteratively to further improve the performance. After obtaining the extrinsic mean $\bar E_c^2[i]$ and variance $\bar F_c^2[i]$, $i = 1,2, \cdots ,D,c = 1,2, \cdots ,\frac{{MNJ}}{K}$ from the second RRH, the {\em a priori} probability is updated at the first RRH as
\begin{align}\label{Decentra_prior}
{P_D}\left( {{{\bf{x}}_c} = {{\bm{\chi }}_j}} \right) \propto & \prod\limits_{i = 1}^D {\exp \left( { - \frac{{{{\left| {{\chi _j}[i] - \bar E_c^2[i]} \right|}^2}}}{{\bar F_c^2[i]}}} \right)} ,\;\forall {{\bm{\chi }}_j} \in {{\bar {\mathbb{A}}}_j},\nonumber\\ & j = \left\lceil {\frac{{cK}}{{MN}}} \right\rceil ,c = 1,2, \cdots ,\frac{{MNJ}}{K}.
\end{align}
By applying GAEP algorithm for a certain number ${n_I}$ of iterations, the first RRH projects the output probabilities $P({\bf{x}}_c^1)$ into the {\em a posteriori} Gaussian distribution ${\cal C}{\cal N}\left( {E_c^{1}[i],F_c^{1}[i]} \right)$, $i = 1,2, \cdots ,D,c = 1,2, \cdots ,\frac{{MNJ}}{K}$ as (\ref{gaussin_appro}).

The extrinsic mean $\bar E_c^1[i]$ and variance $\bar F_c^1[i]$ can be calculated as follows and then delivered to the second RRH,
\begin{subequations}\label{Decentra_extrinc}
\begin{align}
\bar F_c^1[i] = {\left[ {{{\left( {F_c^1[i]} \right)}^{ - 1}} - {{\left( {\bar F_c^2[i]} \right)}^{ - 1}}} \right]^{ - 1}},
\end{align}
\begin{align}
\bar E_c^1[i] = \bar F_c^1[i]\left[ {\frac{{E_c^1[i]}}{{F_c^1[i]}} - \frac{{\bar E_c^2[i]}}{{\bar F_c^2[i]}}} \right].
\end{align}
\end{subequations}

Similarly, the second RRH first updates the {\em a priori} probability ${P_D}\left( {{{\bf{x}}_c}} \right)$, and then produces the {\em a posteriori} Gaussian distribution for each symbol by using GAEP algorithm with $n_I$ iterations. The extrinsic mean ${\bar E_c^2[i]}$ and variance ${\bar F_c^2[i]}$, $i = 1,2, \cdots ,D,c = 1,2, \cdots ,\frac{{MNJ}}{K}$ are finally generated and passed back to the first RRH to form the iterative loop. After a certain number ${n_o}$ of iterations, each RRH obtains a final decision of ${{\bf{\hat x}}}$ in the last iteration step. A detailed implementation of the decentralized GAEP detector is summarized in \textbf{Algorithm \ref{alg:B}}.
\begin{algorithm}
\caption{Decentralized GAEP Detector}
\label{alg:B}
\begin{algorithmic}
\STATE {Input: ${{{\bf{\bar y}}}_1}$, ${{{\bf{\bar y}}}_2}$, ${{\bf{H}}_1}$, ${{\bf{H}}_2}$, ${n_I}$ and ${n_o}$.}
\STATE {Initialization: ${{{\bf{\bar E}}}^u}$ and ${{{\bf{\bar F}}}^u}$, $u = \{ 1,2\} $.}
\FOR{$\mathcal{T} = 0,1, \cdots ,n_o$}
\FOR[parallel computing]{$u = 1,2$}
\STATE {1)\; Update ${P_D}\left( {\bf{x}} \right)$ as in (\ref{Decentra_prior});}
\STATE {2)\; Obtain the output probabilities $P({{\bf{x}}^u})$ by employing GAEP algorithm with ${n_I}$ iterations;}
\STATE {3)\; Project $P({\bf{x}}_c^u)$ into Gaussian distribution ${\cal C}{\cal N}\left( {E_c^{u}[i],F_c^{u}[i]} \right)$, $i = 1,2, \cdots ,D,c = 1,2, \cdots ,\frac{{MNJ}}{K}$;}
\STATE {4)\; Compute the extrinsic mean $\bar E_c^u[i]$ and variance $\bar F_c^u[i]$, $i = 1,2, \cdots ,D,c = 1,2, \cdots ,\frac{{MNJ}}{K}$ as in (\ref{Decentra_extrinc});}
\ENDFOR
\STATE {Exchange the extrinsic mean ${{{\bf{\bar E}}}^u}$ and variance ${{{\bf{\bar F}}}^u}$ between the two RRHs.}
\ENDFOR
\STATE {Output: $P({{\bf{x}}^u}),u = \{ 1,2\}$.}
\end{algorithmic}
\end{algorithm}

\subsection{Discussion}

We can observe that both centralized and decentralized detectors are efficient to recover the signal of each user. The centralized detector requires the RRHs to forward their receptions to the BBU and the decentralized detector requires frequent communications between the RRHs. We now discuss in details about their advantages and disadvantages, respectively. 

From the algorithm discussion, the complexity of the proposed centralized and decentralized detectors are mainly determined by the GAEP algorithm. For each main loop iteration of the GAEP algorithm, (\ref{GAEP_y_m})-(\ref{GAEP_x}) have a complexity order 
$\mathcal{O}(\bar SD)$, $\mathcal{O}(\bar SD)$, $\mathcal{O}(\bar SDQ)$, $\mathcal{O}\left( {\frac{{2MNJDQ}}{K}} \right)$, $\mathcal{O}(2\bar SD)$ and $\mathcal{O}(2\bar SD)$, respectively. Here, we write $\sum\limits_{d = 1}^{2MN} {\left| {\mathcal{I}(d)} \right|}  = \sum\limits_{c = 1}^{{{MNJ} \mathord{\left/
 {\vphantom {{MNJ} K}} \right.
 \kern-\nulldelimiterspace} K}} {\left| {\mathcal{J}(c)} \right|}  = \bar S$ for conciseness. Therefore, the overall complexity orders are ${n_c}\mathcal{O}\left( {6\bar SD + \bar SDQ + \frac{{2MNJDQ}}{K}} \right)$ and ${n_o}{n_I}\mathcal{O}\left( {6\bar SD + \bar SDQ + \frac{{4MNJDQ}}{K}} \right)$ for centralized and decentralized detectors, respectively.\footnote{Note that our proposed centralized and decentralized GAEP detectors can also be applied for OFDM-SCMA counterpart. Similar computational complexity are required for the OFDM-SCMA scenario compared to those of OBSCMA scenario, where ${\bar S}$ represents the approximate numbers of non-zero elements of the channel matrix in OFDM-SCMA scenario.} From our simulations, we observe that the similar values of ${n_c}$ and ${n_o}{n_I}$ are required to guarantee the algorithm convergence. Hence, our proposed centralized and decentralized detectors have comparable computational complexity.

However, the centralized detector may suffer from high communication overhead for information transfer between the RRHs and the BBU, especially when each RRH is with a large number of antennas. The amount of complex-valued data passed from each RRH to the BBU contains $MN{N_u}$ receptions and $3{N_u}{L_u}$ channel state information, where ${N_u}$ represents the number of antennas at $u$-th RRH and ${L_u} = \sum\limits_{j = 1}^J {{L_{uj}}}, \forall u = \{ 1,2\}$. The BBU then broadcasts ${\frac{{MNJD}}{K}}$ detected complex-valued data to each RRH afterwards. Therefore, the overall complex-valued data passed between the RRHs and the BBU in centralized detector is $MN({N_1} + {N_2}) + 3({N_1}{L_1} + {N_2}{L_2}) + \frac{{2MNJD}}{K}$. For the decentralized detector, the RRHs execute local computing processing parallelly, and coordinate with each other with limited consensus information exchange. The exchanged information only includes $\frac{{MNJD}}{K}$ means and $\frac{{MNJD}}{K}$ variances in each iteration, resulting in $\frac{{2MNJD}}{K}{n_o}$ complex-valued data transferred among the RRHs overall. Such a small amount of data exchange can not only alleviate the excessively requirement on interconnection bandwidth among the decentralized RRHs, but also avoid the large data transfer between the RRHs and the BBU in the centralized detector.

In addition, the BBU is generally located far away from the RRHs, and requires a high computing capacity to solve the large dimension problem of multi-user detection. Therefore, the centralized detector results in high-latency communications and effects on user experience. Nevertheless, the computations can be carried out in a decentralized and parallel fashion between the two neighbouring RRHs in the decentralized detector, which leads to a relatively low-latency communication system.

Note that our decentralized detector is not a substitute for the centralized detector but a powerful complement. It enables processing at the edges (i.e., RRHs) while still offering the possibility to process at the cloud (i.e., BBU) if the backhaul bandwidth and computation are sufficient to support latency-tolerant applications. To summarize, we include the detailed comparisons of the centralized and decentralized detectors in TABLE~\ref{tab1}.
\begin{table*}
  \begin{center}
    \caption{Detailed comparisons of the centralized and decentralized detectors.}
    \label{tab1}
    \begin{tabular}{c|c|c|c} 
    \hline
      Detectors & Computational Complexity & Communication Overhead & Latency\\
      \hline
      Centralized detector & ${n_c}\mathcal{O}\left( {6\bar SD + \bar SDQ + \frac{{2MNJDQ}}{K}} \right)$ & $MN({N_1} + {N_2}) + 3({N_1}{L_1} + {N_2}{L_2}) + \frac{{2MNJD}}{K}$ & High\\
      Decentralized detector & ${n_o}{n_I}\mathcal{O}\left( {6\bar SD + \bar SDQ + \frac{{4MNJDQ}}{K}} \right)$ & $\frac{{2MNJD}}{K}{n_o}$ & Low\\
      \hline
    \end{tabular}
  \end{center}
\end{table*}


\section{Simulation Results}\label{V_simulation}

In this section, we evaluate the performance of our proposed centralized and decentralized detectors for OBSCMA with CoMP systems. In our simulations, the carrier frequency is centered at 4 GHz and subcarrier spacing $\Delta f = 15$ kHz. The rolloff factor of the RRC filters is set to 0.4 for both the transmitter and receiver. Unless otherwise specified, we consider a delay-Doppler plane with $M = 64$ and $N = 16$. We also assume that $J = 6$ users are sharing $K = 4$ orthogonal resources simultaneously, and set $D = 2$ non-zero entries in each codeword. The user-specific codebooks are designed according to \cite{xiao2018capacity} with size $Q = 4$, and the transmission power of each mobile user is assumed to be the same. 

As shown in Fig. \ref{fig1_CoMP}(a), we consider a scenario where the mobile users are uniformly and independently distributed in the cell. The RRHs are positioned along the highway, spacing each other by ${d_h} = 1000$ m. The perpendicular distance of the RRHs and the highway line ${d_p} = 150$ m, and the width distance of the highway road ${d_w} = 50$ m. The distance-dependent pathloss propagation is modeled as $PL(d)[\text{dB}] = 142.1 + 37.6{\log _{10}}(d)$ \cite{tao2016content}, where $d$ is the distance in kilometers. The noise power spectral density is set to be -174 dBm/Hz for both RRHs. We adopt a typical urban channel model \cite{failli1989digital} with exponential power delay profile. The velocity of the $j$-th mobile user is set to 300 km/h, leading to a maximum Doppler frequency shift ${\nu _{j,\max }} = 1111$ Hz,  $\forall j = \{ 1,2, \cdots ,J\} $. We further generate the Doppler frequency shift for the $i$-th delay of the channel between the $j$-th user and $u$-th RRH by using the Jakes formulation \cite{ge2021receiver,raviteja2018interference}, i.e., ${\nu _{uj,i}} = {\nu _{j,\max }}\cos \left( {{\rho _{uj,i}}} \right)$, where ${{\rho _{uj,i}}}$ is uniformly distributed over $\left[ {\frac{\pi }{2},\pi } \right]$ if the $j$-th user is moving away from the $u$-th RRH, and distributed over $\left[ {0,\frac{\pi }{2}} \right]$ if the $j$-th user is moving towards the $u$-th RRH.

We assume that the full CSIs are known at the receiver and also test the effect of channel uncertainties on receiver performance. After extensive experimentations, we select $\Delta  = 0.3$, $\varepsilon ={10^{ - 8}}$, $\varrho=0.1$ and ${n_c}=20$ to yield attractive compromise between convergence speed and accuracy. The simulation results are average over 1000 independent Monte Carlo trails.

\begin{figure}
  \centering
  \includegraphics[width=3.6in]{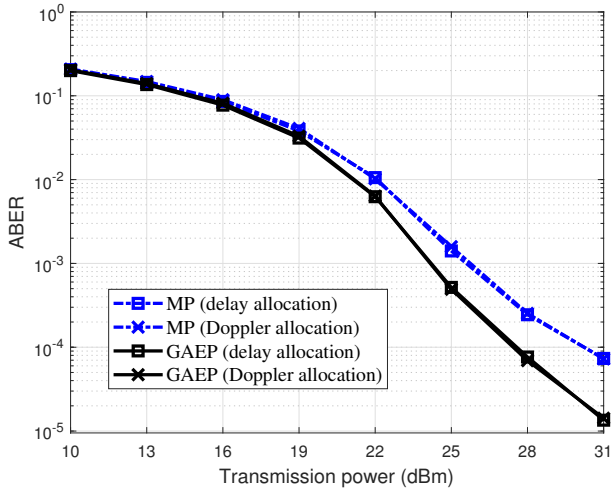}
  \caption{ABER performance comparison for different SCMA codewords allocations and detection algorithms.}\label{Simu1_allocation}
\end{figure}
We first investigate the effects of SCMA codewords allocation on receiver performance.
Fig. \ref{Simu1_allocation} illustrates the ABER performance of the proposed OBSCMA with CoMP system for different SCMA codewords allocations. Without loss of generality, we apply centralized GAEP detector and allocate the SCMA codewords either along the delay axis in Fig. \ref{fig2_allocation}(a) (denoted as delay allocation) or along the Doppler axis in Fig. \ref{fig2_allocation}(b) (denoted as Doppler allocation). To highlight the superiority of the proposed GAEP algorithm, we also provide the baseline performance of traditional MP algorithm \cite{mishra2021otfs,raviteja2018interference} in Fig. \ref{Simu1_allocation}. We can observe that the proposed GAEP algorithm outperforms the MP algorithm in the considered OBSCMA with CoMP system. We also notice that the SCMA codewords allocation has slightly effect on the receiver performance. 
In the rest of our simulations, we shall allocate the SCMA codewords along the delay axis unless otherwise noted. 

\begin{figure}
  \centering
  \includegraphics[width=3.6in]{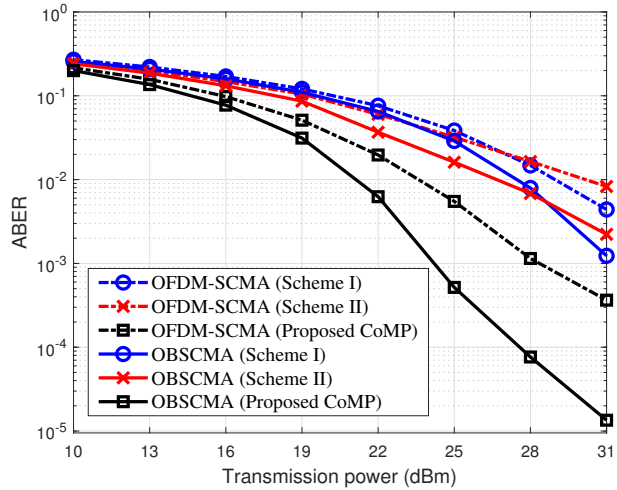}
  \caption{ABER performance comparison for different scheme scenarios.}\label{Simu2_compare}
\end{figure}
Fig. \ref{Simu2_compare} compares the ABER performance of the proposed scheme with those of co-located RRHs scheme shown in Fig. \ref{fig1_CoMP}(b) and traditional cellular network scheme shown in Fig. \ref{fig1_CoMP}(c), denoted as Scheme I and Scheme II, respectively.
We also produce the performance of traditional OFDM-SCMA counterparts as benchmarks for different schemes in Fig. \ref{Simu2_compare}. Note that the proposed GAEP algorithm can be generalized to the OFDM-SCMA scenarios in a straightforward manner, thus, we omit the details here for brevity. The results reveal that all the receivers benefit from higher transmission power. The OBSCMA can achieve better performance than its OFDM-SCMA counterparts for each scheme. Our proposed OBSCMA with CoMP system outperforms the other schemes significantly owing to the utilization of sufficient underlying channel diversity.

We also note that as transmission power increases, the ABER performance of Scheme I and Scheme II intersect with each other for both the OBSCMA and OFDM-SCMA scenarios. This is due to the fact that Scheme II experiences favorable propagation gain but limits to the spatial diversity gain.

\begin{figure}
  \centering
  \includegraphics[width=3.6in]{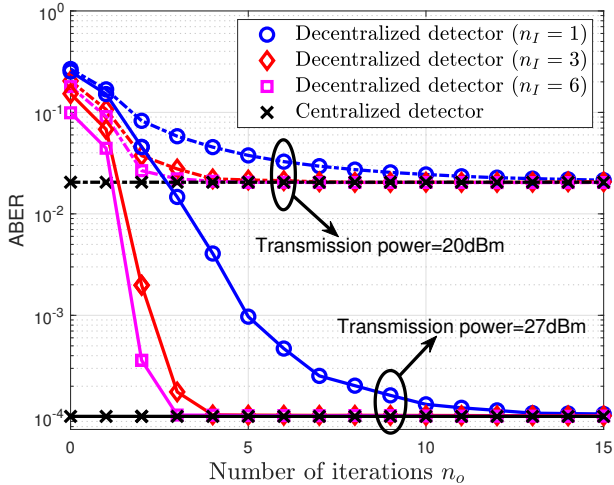}
  \caption{ABER performance comparison of the proposed centralized and decentralized detectors.}\label{Simu3_iteration}
\end{figure}

Fig. \ref{Simu3_iteration} shows the ABER performance of our proposed centralized and decentralized detectors for OBSCMA with CoMP system. The results clearly show that the ABER performance of decentralized detector with sufficient number of iterations would asymptotically approach that of centralized detector. To guarantee the convergence of decentralized detector, a larger iteration number $n_I$ requires a relatively smaller value of $n_o$, and vice versa, i.e., a smaller number $n_I$ demands a larger value of $n_o$. Therefore, our proposed decentralized detector can yield desirable trade-off between the local processing efficiency of each RRH and the communication overhead between the RRHs.

\begin{figure}
  \centering
  \includegraphics[width=3.6in]{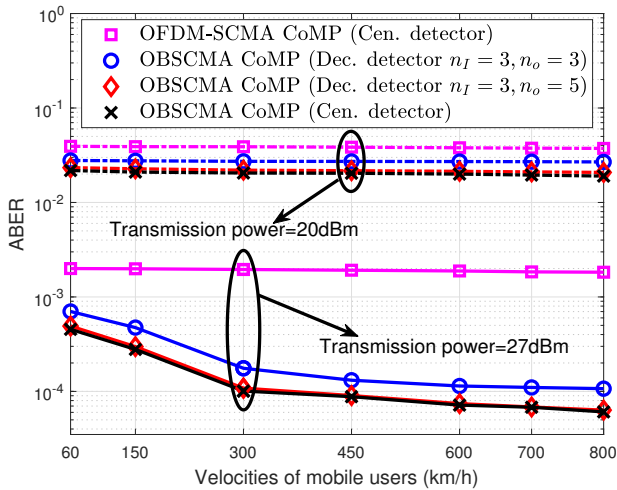}
  \caption{ABER performance of centralized and decentralized detectors with different user mobile velocities.}\label{Simu4_velocity}
\end{figure}

In Fig. \ref{Simu4_velocity}, we test the ABER performance of the proposed centralized and decentralized detectors for OBSCMA with CoMP system under different user mobile velocities. We also present the performance of centralized detector for OFDM-SCMA counterpart as benchmark. As the velocities of the mobile users grow, the ABER performance of our centralized and decentralized detectors for OBSCMA with CoMP system improves gradually and saturates after velocity beyond $600$ km/h, particularly for high transmission power. This is attributed to the reason that OTFS modulation can resolve more distinct paths in the Doppler domain with the help of higher user velocity. As a result, better performance becomes possible. However, the performance improvement of OFDM-SCMA counterpart is negligible as the OFDM modulation cannot efficiently exploit the Doppler diversity. We again notice that the ABER performance of decentralized detector approach that of centralized one for our OBSCMA with CoMP system when the iteration numbers are sufficient, and slightly degrade when the iteration numbers are inadequate for different velocities.

\begin{figure}
  \centering
  \includegraphics[width=3.6in]{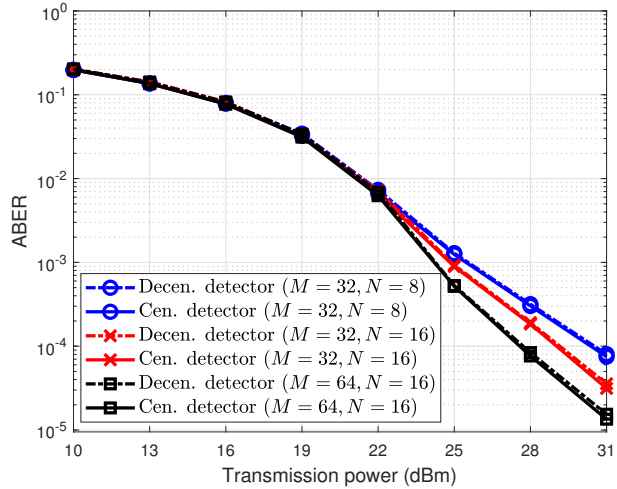}
  \caption{ABER performance comparison for different system settings of $M$ and $N$ ($n_I=3$ and $n_o=5$).}\label{Simu5_MN}
\end{figure}

Fig. \ref{Simu5_MN} further illustrates the ABER performance for both the centralized and decentralized detectors (with $n_I=3$ and $n_o=5$) under different system settings of $M$ and $N$. We notice that the ABER performance of both centralized and decentralized detectors improve as $M$ and $N$ increase due to the higher resolution of OTFS delay-Doppler grid. This leads to the diversity benefit as the receiver can resolve a larger number of signal paths in the channel.

\begin{figure}
  \centering
  \includegraphics[width=3.6in]{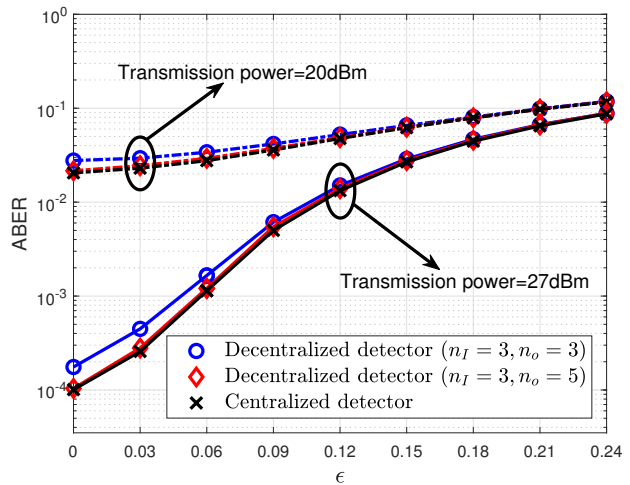}
  \caption{ABER performance of centralized and decentralized detectors with imperfect CSIs.}\label{Simu6_CSI}
\end{figure}

Finally, we test the effect of imperfect CSIs on the ABER performance of the proposed centralized and decentralized detectors in Fig. \ref{Simu6_CSI}. Here, we capture the imperfect information of CSIs by adopting the following norm-bounded uncertainty model \cite{ge2021receiver}:
\begin{align*}
{h_{uj,i}} = {{\hat h}_{uj,i}} + \Delta {h_{uj,i}},\;\left\| {\Delta {h_{uj,i}}} \right\| \le {\epsilon _{{h_{uj,i}}}},\\
{\tau _{uj,i}} = {{\hat \tau }_{uj,i}} + \Delta {\tau _{uj,i}},\;\left\| {\Delta {\tau _{uj,i}}} \right\| \le {\epsilon _{{\tau _{uj,i}}}},\\
{\nu _{uj,i}} = {{\hat \nu }_{uj,i}} + \Delta {\nu _{uj,i}},\;\left\| {\Delta {\nu _{uj,i}}} \right\| \le {\epsilon _{{\nu _{uj,i}}}},
\end{align*}
where ${{\hat h}_{uj,i}}$, ${{\hat \tau }_{uj,i}}$ and ${{\hat \nu }_{uj,i}}$ are the estimated values of ${h_{uj,i}}$, ${\tau _{uj,i}}$ and ${\nu _{uj,i}}$, respectively. $\Delta {h_{uj,i}}$, $\Delta {\tau _{uj,i}}$ and $\Delta {\nu _{uj,i}}$ represent the corresponding channel estimation errors, which are norm-bounded by the given radius ${\epsilon _{{h_{uj,i}}}}$, ${\epsilon _{{\tau _{uj,i}}}}$ and ${\epsilon _{{\nu _{uj,i}}}}$, respectively. For simplicity, we assume that ${\epsilon _{{h_{uj,i}}}} = \epsilon \left\| {{{\hat h}_{uj,i}}} \right\|$, ${\epsilon _{{\tau _{uj,i}}}} = \epsilon \left\| {{{\hat \tau }_{uj,i}}} \right\|$ and ${\epsilon _{{\nu _{uj,i}}}} = \epsilon \left\| {{{\hat \nu }_{uj,i}}} \right\|$, $\forall u,j,i$. From the results in Fig. \ref{Simu6_CSI}, we observe that the performance degradation of our proposed centralized and decentralized detectors is mild for modest levels of channel uncertainty $\epsilon$. As the receiver performance gradually drops in accordance with the levels of channel uncertainty, our proposed schemes and detectors exhibit robustness against typical CSI errors.

\section{Conclusion}\label{VI_conclusion}
In this paper, we proposed an OBSCMA with CoMP system, where a group of SCMA mobile users modulated in accordance with OTFS, and served by the RRH in front of them and the RRH behind them at the same time. The proposed OBSCMA with CoMP framework can efficiently support massive mobile connectivity and naturally harvest sufficient diversity for better performance. Based on the ML detector, we analyzed the single-user ABER bound as the benchmark of our OBSCMA with CoMP system. We also developed practical receiver algorithms for centralized and decentralized detectors to exploit the underlying channel diversity from the receptions of the RRHs. Our results demonstrated the superiority of the proposed OBSCMA with CoMP system compared to existing solutions, such as co-located RRHs and traditional cellular networks, as well as their OFDM-SCMA counterparts. The proposed centralized and decentralized detectors are both effectiveness and robustness against CSI uncertainty.
The interesting future works include the coordinated design for more densely deployment of RRHs, the development of fast handover and efficient resource management, efficient robust receiver design based on realistic channels, grant-free random access for massive machine-type vehicle connectivity and so on.


%





\ifCLASSOPTIONcaptionsoff
  \newpage
\fi



%
%




\bibliographystyle{IEEEtran}
\footnotesize
\bibliography{ref_OTFS_CoMP}

%

\begin{IEEEbiography}[{\includegraphics[width=1in,height=1.25in,clip,keepaspectratio]{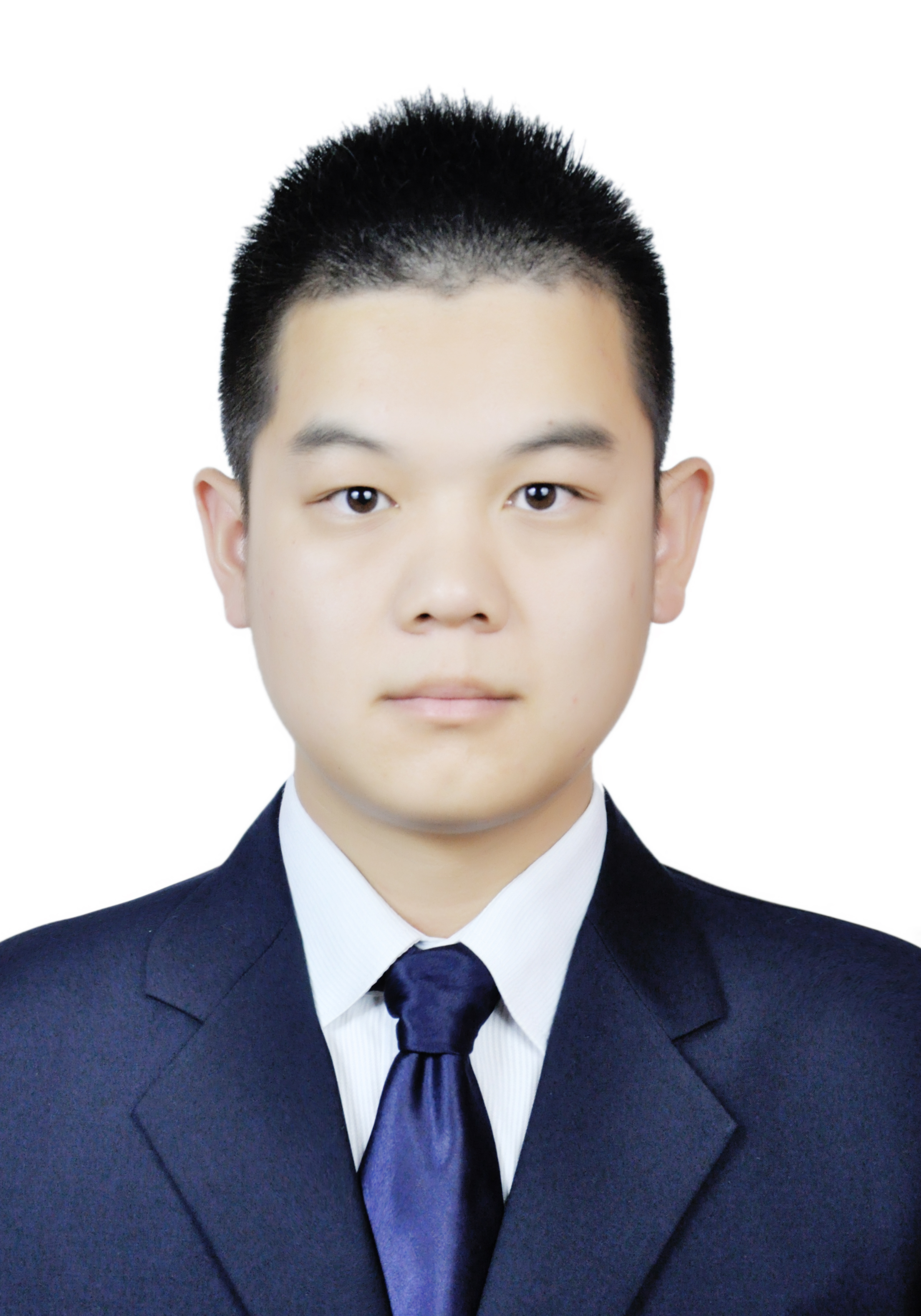}}]{Yao Ge}(Member, IEEE) received the Ph.D. degree in Electronic Engineering from The Chinese University of Hong Kong (CUHK), Shatin, Hong Kong, in 2021, and the M.Eng. degree (research) in Communication and Information System and the B.Eng. degree in Electronics and Information Engineering from Northwestern Polytechnical University (NPU), Xi’an, China, in 2016 and 2013, respectively.
He is currently a Research Fellow with Continental-NTU Corporate Lab, Nanyang Technological University (NTU), Singapore. From October 2015 to March 2016, he was a Visiting Scholar with the Department of Electrical and Computer Systems Engineering, Monash University, Melbourne, VIC, Australia. From April 2016 to August 2016, he was a Visiting Scholar with the Department of Computer, Electrical and Mathematical Science and Engineering, King Abdullah University of Science and Technology (KAUST), Thuwal, Saudi Arabia. From May 2019 to December 2019, he was a Visiting Scholar with the Department of Electrical and Computer Engineering, University of California at Davis (UC Davis), Davis, CA, USA.
His current research interests include wireless communications and system design, Internet of Things, cognitive radio networks, automotive vehicle signal processing and communications, wireless network security, statistical signal processing, optimization and game theory. He received the Best Paper Award from the International Conference on Wireless Communications and Signal Processing (WCSP) 2022. He is the Founding Member of the IEEE ComSoc Special Interest Group (SIG) on OTFS and has served as the Co-Chair for the IEEE/CIC ICCC 2022 Workshop on OTFS.
\end{IEEEbiography}

\begin{IEEEbiography}[{\includegraphics[width=1in,height=1.25in,clip,keepaspectratio]{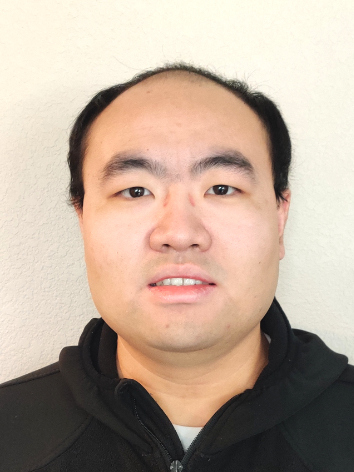}}]{Qinwen Deng}(Student Member, IEEE) received the B.Eng. degree in Electronic Information Engineering from the University of Science and Technology of China (USTC), Hefei, China, in 2018. He is currently pursuing the Ph.D. degree with the Department of Electrical and Computer Engineering, University of California at Davis (UC Davis), Davis, CA, USA.
His current research interests include wireless communications and hypergraph signal processing.
\end{IEEEbiography}

\begin{IEEEbiography}[{\includegraphics[width=1in,height=1.25in,clip,keepaspectratio]{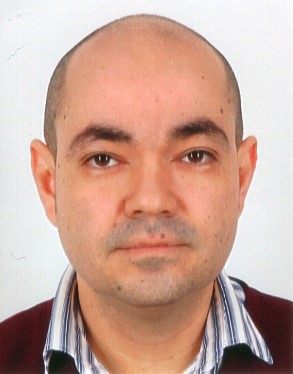}}]{David~Gonz\'{a}lez~G.}(Senior Member, IEEE) received the master's degree in mobile communications and the Ph.D. degree in signal theory and communications from the Universitat Politècnica de Catalunya, Spain. He has served as a Postdoctoral Fellow with the Department of Communications and Networking, Aalto University, Finland, from 2014 to 2017. From 2017 to 2018, he served as a Research Engineer with the Panasonic Research and Development Center, Germany. Since 2018, he has been with Continental AG, Germany. He has represented his previous and current company as delegate in the 3GPP for 5G standardization, mainly focused on physical layer aspects and vehicular communications. His research is focused on diverse aspects of cellular networks and wireless communications, including interference modeling and coordination, radio access modeling and optimization, resource allocation, and vehicular communications (V2X). 
\end{IEEEbiography}

\begin{IEEEbiography}[{\includegraphics[width=1in,height=1.25in,clip,keepaspectratio]{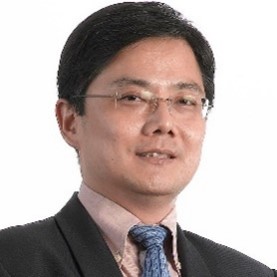}}]{Yong Liang Guan}(Senior Member, IEEE) obtained his Ph.D. degree from the Imperial College London, London, U.K., and Bachelor of Engineering with first class honours from the National University of Singapore, Singapore. He is currently a Professor of Communication Engineering with the School of Electrical and Electronic Engineering, Nanyang Technological University (NTU), Singapore, where he leads the Continental-NTU Corporate Lab and the successful deployment of the campus-wide NTU-NXP V2X Test Bed. His research interests broadly include coding and signal processing for communication systems and data storage systems. He is an Editor of the {\sc IEEE Transactions on Vehicular Technology}. He is also an Associate Vice President of NTU and a Distinguished Lecturer of the IEEE Vehicular Technology Society (2021-2023).
\end{IEEEbiography}

\begin{IEEEbiography}[{\includegraphics[width=1in,height=1.25in,clip,keepaspectratio]{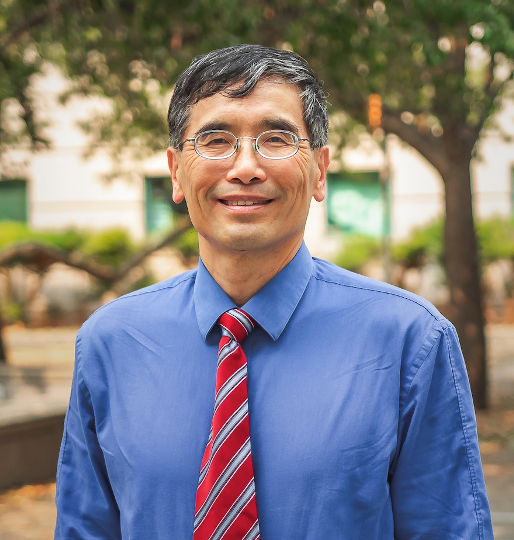}}]{Zhi Ding}(Fellow, IEEE) is with the Department of Electrical and Computer Engineering at the University of California at Davis (UC Davis), Davis, where he currently holds the position of distinguished professor. He received his Ph.D. degree in Electrical Engineering from Cornell University in 1990. From 1990 to 2000, he was a faculty member of Auburn University and later at The University of Iowa. Prof. Ding joined the College of Engineering at UC Davis in 2000. His major research interests and expertise cover the areas of wireless networking, communications, signal processing, multimedia, and learning. Prof. Ding supervised over 30 PhD dissertations since joining UC Davis. His research team of enthusiastic researchers works very closely with industry to solve practical problems and contributes to technological advances. His team has collaborated with researchers around the world and welcomes self-motivated young talents as new members.

Prof. Ding is a Fellow of IEEE and currently serves as the Chief Information
Officer and Chief Marketing Officer of the IEEE Communications Society. He was an associate editor for {\sc IEEE Transactions on Signal Processing} from 1994-1997, 2001-2004, and an associate editor of {\sc IEEE Signal Processing Letters} 2002-2005. He was a member of Technical Committee on Statistical Signal and Array Processing and a member of Technical Committee on Signal Processing for Communications (1994-2003). Prof. Ding was the General Chair of the 2016 IEEE International Conference on Acoustics, Speech, and Signal Processing and the Technical Program Chair of the 2006 IEEE Globecom. He was also an IEEE {\em Distinguished Lecturer} (Circuits and Systems Society, 2004-06, Communications Society, 2008-09). He served on as {\sc IEEE Transactions on Wireless Communications} Steering Committee Member (2007-2009) and its Chair (2009-2010). Prof. Ding is a coauthor of the textbook: {\em Modern Digital and Analog Communication Systems}, 5th edition, Oxford University Press, 2019. Prof. Ding received the IEEE Communication Society’s WTC Award in 2012 and the IEEE Communication Society’s Education Award in 2020. 
\end{IEEEbiography}




\end{document}